\documentstyle[psfig]{mn}
\input epsf

\newcommand{\ebv}{ {\it E(B--V)}}

\newcommand{\bb}{\bibitem[]{bla}}
\newcommand{\zm}{ \relax \ifmmode {\rm M_{\odot}} \else {M$_{\odot}$}\fi}

\newcommand{\ang}{$\rm \AA$}

\newcommand{\degree}{$^{\rm o}$}
\newcommand{\arc}{$^{\prime\prime}$}
\newcommand{\arcm}{$^{\prime}$}
\newcommand{\mic}{$\mu$m}
\newcommand{\ea}{{et al.}}

\newcommand{\km}{km s$^{-1}$}
\newcommand{\ha}{H$\alpha$}

\def\lesssim{\mathrel{\hbox{\rlap{\hbox{\lower4pt\hbox{$\sim$}}}\hbox{$<$}}}}
\def\gtrsim{\mathrel{\hbox{\rlap{\hbox{\lower4pt\hbox{$\sim$}}}\hbox{$>$}}}}

\def\ion#1#2{#1$\;${\small\rm\@Roman{#2}}\relax}
 
\newcommand{\Av}{$A_{V}$}

\newbox\grsign \setbox\grsign=\hbox{$>$} \newdimen\grdimen 
\grdimen=\ht\grsign
\newbox\simlessbox \newbox\simgreatbox
\setbox\simgreatbox=\hbox{\raise.5ex\hbox{$>$}\llap
     {\lower.5ex\hbox{$\sim$}}}\ht1=\grdimen\dp1=0pt
\setbox\simlessbox=\hbox{\raise.5ex\hbox{$<$}\llap
     {\lower.5ex\hbox{$\sim$}}}\ht2=\grdimen\dp2=0pt

\makeatletter 
\renewcommand\@biblabel[1]{}     

\makeatother

\begin{document}

\title[\ha\ spectropolarimetry of  { B[e]}  and Herbig Be stars]
{
\ha\ spectropolarimetry of  B[e]
and Herbig Be stars }

\author[Ren\'e D. Oudmaijer \& Janet E. Drew ]
{ Ren\'e D. Oudmaijer \& Janet E. Drew \\
Imperial College of Science, Technology and Medicine,
Blackett Laboratory, Prince Consort Road,\\ London,  SW7 2BZ, U.K.   
}

\date{received,  accepted}

\maketitle
\begin{abstract}

    We present the results of medium resolution ($\Delta v$ $\approx$  
60 km s$^{-1}$)
spectropolarimetric observations across H$\alpha$ of a sample of B[e] and
Herbig Be objects.  A change in linear polarization across H$\alpha$ is
detected in a large fraction of the objects, with characteristics ranging
from simple depolarization in a couple of Herbig Be stars, to more
complex behaviour in the probable post main sequence B[e] stars.  H$\alpha$
in the spectra of HD 37806 and HD 50138 each consist of a double-peaked
polarized line and a superposed unpolarized single emission peak, suggesting
two distinct line-forming regions.  Multiple observations of HD 45677 allow
for the separation of electron and dust scattering effects for the first
time: the difference between derived intrinsic polarization angles of the 
two components indicate that the dust-scattering region is clumpy.  
Two unexpected results are the non-detections of H$\alpha$ polarization 
changes in $\omega$~Ori, where depolarization has previously been detected,
and in MWC 297, which exhibits source elongation at radio wavelengths.
In $\omega$~Ori time variability is probably responsible such that this 
star's electron-scattering disk was much weakened at the time of 
observation.  Two hypotheses are advanced that might explain the MWC~297 
result.

    The general findings are that roughly half of the observed Herbig Be 
stars show polarization changes across H$\alpha$, implying immediately that 
their ionized envelopes are not spherically symmetric.  This pattern, if
confirmed by observations of a larger sample, could indicate that the
non-detection rate is simply a consequence of sampling randomly-oriented
circumstellar disks able to scatter starlight within a few stellar radii.
The stars classified as B[e] stars all show startling polarization changes
across H$\alpha$.  The details in each case are different, but the widely
accepted concept of dense H$\alpha$-emitting equatorial disks around these
objects is supported.

\end{abstract}


\section{Introduction}

   Classification of a star as a Be star has long been recognised as
consignment to a loosely defined phenomenological group, rather than
as a definition of the evolutionary status of the star.  Presently the
Be stars may be divided into three main groups: (i) the classical Be
stars, generally thought of as the most rapidly rotating near main
sequence B stars (see e.g. Slettebak, 1988); (ii) the Herbig Be stars,
first identified by Herbig (1960) as stars in the B spectral type
range whose association with star-forming regions and emission line
character might indicate that they are very young; (iii) the B[e]
stars (Allen \& Swings, 1976; Zickgraf \ea \ 1985), noted for
the presence of forbidden line as well as H{\sc i} emission in their
spectra and strong IR continuum excesses (also seen in Herbig Be
stars).  While it was initially thought that B[e] stars are
preferentially supergiants, recent work (Gummersbach et al. 1995) has
demonstrated from deep LMC observations that B[e] characteristics may
also be seen at significantly lower luminosities.

   It is now widely accepted that classical Be stars are encircled at
their equators by ionized, low opening angle, almost Keplerian, disks,
that are optically-thick in H$\alpha$.  By contrast, the circumstellar
geometry of Herbig Be and B[e] stars is rather more of an open
question.  In this work, we will open up another avenue for exploring
this issue.  We present medium resolution spectropolarimetry across
the \ha\ line, a technique that, in the case of the detection of
polarization changes across the line can provide an  answer to
the most basic question ``Is the ionized material around these stars
spherically symmetric or not?''.

   By comparing \ha\ polarization with that of the continuum one can
exploit the fact that line and continuum respectively form within a
larger and smaller volume and subsequently `see' different scattering
geometries.  Essentially, \ha\ is not significantly scattered by the
ionized envelope in which it forms, whereas the continuum arising
primarily from the central star embedded in the envelope undergoes
electron scattering.  In the case that the ionized envelope's
projection on to the plane of the sky is non-circular, a net linear
polarization is imprinted on the continuum light, but not on \ha\,
producing a drop in the polarization percentage across the line
(`line-effect').  The addition of further continuum
polarization by either a dusty envelope or the ISM modifies this
change and may even produce a net percentage rise across the line --
but, significantly, it cannot nullify the change.  For example,
Schulte-Ladbeck \ea\ (1994) showed that the \ha \ emission line of AG
Car displayed enhanced polarization at one epoch while on other
occasions a de-polarization across the line was observed. After the
correction for ISP however, \ha \ was de-polarized with respect to the
continuum on all occasions.

The advantage of spectropolarimetry over broadband polarimetry is that
a result can be obtained even where it is not possible to distinguish
the various contributions to the total continuum polarization.
Furthermore, by spectrally-resolving the H$\alpha$ line profile one
can hope to pick out more subtle effects arising in cases where the
assumption that the \ha\ emission is unscattered, and hence
unpolarized, breaks down.  Qualitatively these were demonstrated in
model calculations by Wood, Brown \& Fox (1993).  When there is
significant scattering of \ha\, the line profile in linear polarized
light becomes a probe of the velocity field in the electron-scattering
medium.  Using this tool we have already shown in the case of the B[e]
star, HD 87643, that there is direct evidence of a rotating and
expanding outflow (Oudmaijer et al 1998).

    The pioneering work in this area was made in the seventies, when Clarke 
\& McLean (1974), Poeckert (1975) and Poeckert
\& Marlborough (1976, hereafter PM) conducted narrow-band polarimetric
studies of Be stars that compared the linear polarization on and off
\ha . Many instances of line de-polarization were found, showing that the 
envelopes of Be stars do not project as circles onto the sky. After this 
time, polarimetric studies were made of several classes of object, 
but due to observational difficulties it remained a specialist activity.
However in the past few years there has been rising interest in the technique.
Spectropolarimetry has been performed on several strong \ha\ emitting evolved 
stars, such as AG Car and HR Car (Schulte-Ladbeck \ea\ 1994, Clampin \ea\ 
1995), where the position angle of the spatially-unresolved flattened 
electron scattering region has been shown to agree with the observed 
extension of the optically visible 
nebulae surrounding these objects.  Both the B[e] and Herbig Be stars are
ideal objects to subject to this style of observation, since they are 
strong \ha\ emitters and often optically bright enough to render studies at 
medium resolution with high photon counts feasible with 4-meter class 
telescopes.  Furthermore there is a clear need for this type of observation
since a change in the linear polarization across \ha\ can be the only 
direct evidence of electron scattering operating on the scale of a few 
stellar radii as opposed to polarization by a dusty envelope.

     In the first instance, the observations presented here were
motivated by the aim of examining Herbig Be stars for the presence of
ionized circumstellar disks.  These reputedly intermediate mass
objects present a phenomenology that suggests they are approaching or
have recently achieved a main sequence location on the HR diagram --
they are the higher mass counterparts of the T Tauri stars.  The
paradigm for star formation invokes a collapsing cloud and
conservation of angular momentum that results in the formation of a
flattened circumstellar (accretion) disk, that eventually accretes or
is blown away by an outflow.  However there is not yet a consensus
that accretion disks are commonly associated with the known Herbig Be
(and Ae) stars.  There is a certain irony that T Tauri stars, their
lower mass counterparts, are generally accepted to have disk-like
envelopes (e.g. HH30 in Burrows et al. 1996), while evidence is
accumulating that their higher mass counterparts, the optically
obscured massive Young Stellar Objects (YSOs) are also surrounded by
disk-like structures (e.g. Hoare \& Garrington 1995, and references
therein).

 MERLIN radio data on MWC 297, a nearby radio-bright early Herbig Be
star, also reveals an elongated (but ionized) structure on a spatial
scale of $\sim$100~AU (Drew \ea\ 1997).  More direct high resolution
imaging is clearly worthwhile and of course spectropolarimetry can
help identify interesting targets.  Nevertheless, at the present,
there persists a debate that the observed spectral energy
distributions  require, on the one hand, dusty disks
(e.g. Malfait, Bogaert \& Waelkens, 1998) or, on the other, that they
can be fit satisfactorily by  spherically symmetric dusty
envelopes (Miroshnichenko, Ivezi\'c \& Elitzur, 1997; see also the
overview of this issue in Pezzuto, Strafella \& Lorenzetti 1997).
The recent direct detection of a rotating disk around a Herbig Ae
star by Mannings, Koerner \& Sargent (1997) indicates that at least
some of these objects have disk-like geometries.  Broad-band
polarimetry of a number of Herbig stars has revealed variability of
the polarization of the objects which could imply deviations from
spherical symmetry of the dusty envelopes (e.g. Grinin \ea\ 1994, who
studied UX Ori; Jain \& Bhatt, 1995). By contrast, the \ha \
spectropolarimetry traces scales even closer to the star, the ionized
material.

   With regard to the B[e] stars, first picked out by Allen \&
Swings (1976), the argument for embedding them in disk-like equatorial
structures has largely been won in that there is widespread acceptance
of Zickgraf's phenomenological model (Zickgraf et al 1985, 1986).
This is because there is compelling spectral evidence of a fast,
presumably polar, wind at UV wavelengths, that combines with a high
emission measure, much more slowly expanding, presumably equatorial,
flow traced by strong optical emission lines.  Broad-band polarimetry
by Zickgraf \& Schulte-Ladbeck (1989) and Magalh$\tilde{\rm a}$es
(1992) indicate that for a sub-sample of B[e] objects, the
circumstellar dust, located at larger distances from the star, is
distributed in a geometry deviating from spherically symmetric.  The
unresolved issue is how these axially-symmetric structures arise and
indeed what the stellar evolutionary status of this object class
really is.  The fact that B[e] stars are far from being exclusively
supergiants deepens the mystery.  In this context, Herbig's (1994)
concern about the difficulty of distinguishing Herbig Be from B[e]
stars becomes all the more intriguing.  To progress in understanding
how B[e] disks arise, a more complete description of the disk density
and velocity field is highly desirable.  It is in this respect that
\ha\ spectropolarimetry has the potential to provide unique insights.

     Because of the problems of distinguishing between the B[e] and
Herbig Be categories, there is always a significant probability that a
Herbig Be sample contains some B[e] stars. Indeed, for Galactic B[e]
stars it is often difficult to determine whether an object is a
luminous evolved object or a less luminous pre-main sequence object
(see e.g. the discussions on HD 87643; Oudmaijer \ea 1998, MWC 137;
Esteban \& Fern\'andez 1998, and HD 45677; de Winter \& van den Ancker
1997).  Here we exploit this in that our programme of \ha\
spectropolarimetry programme includes as targets relatively clear-cut
examples of post main sequence B[e] stars alongside undisputed Herbig
Be stars and objects that might be either.  In this paper we give an
overview of our observing campaign to date.  In Section~2, the way in
which targets were selected and the observations are discussed.  The
results and their interpretation are presented on a case-by-case basis
in Sec.~3. Sec.~4 contains a discussion on the power of
spectropolarimetry and what we have learned from this program.  We
conclude in Sec.~5.

\begin{table*}
\caption{Targets 
\label{targets}}
\begin{tabular}{llllllll}
\hline
\hline
Name          & Other name  & {\it V} &Spectral Type  & Date & Integration (s)        \\
\hline
\hline
HD 37806      & MWC 120     &  8.0  & B9Ve+sh  & 11-01-95 & 4$\times$120 \\     
              &             &&                 & 31-12-96 & 16$\times$90 \\           
$\omega$ Ori  & HD 37490    &  4.6  & B3III/IVe& 11-01-95 & 8$\times$15  \\     
V380 Ori      & BD-06 1253  &  10.0 & B8/A1e   & 31-12-96 & 12$\times$300\\       
MWC 137       & PK 195-00.1 &  11.2 & B0ep     & 31-12-96 & 16$\times$200, 8$\times$300\\ 
HD 259431     & MWC 147     &  8.7  & B5Vep    &30-12-96  & 4$\times$240, 8$\times$180\\
HD 45677      & FS CMa      &  7.6  & B3[e]p+sh& 11-01-95 & 24$\times$20 \\     
              &             &&                 &30-12-96  & 32$\times$10 \\          
HD 50138      & MWC 158     &  6.7  & B6V[e]+sh& 11-01-95 & 16$\times$30 \\     
              &             &&                 & 01-01-97 & 24$\times$60\\      
AS 116        & BD-10 1351  &  9.6  & Be       & 01-01-97 & 32$\times$300 \\ 
HD 53367      & MWC 166     &  6.9 & B0III/IVe&30-12-96  & 8$\times$90   \\
              &             &&                & 01-01-97  & 8$\times$240  \\       
Lk\ha\ 218    &             &  11.9 & B6e     & 31-12-96 & 20$\times$300 \\
HD 52721      & GU CMa      &  6.6  & B2Vne    & 11-01-95 & 8$\times$100 \\          
HD 76534      & He 3-225    &  8.0  & B2/3ne   & 11-01-95 & 8$\times$300 \\         
              &             &&                 & 31-12-95 & 4$\times$ 300 \\         
              &             &&                 & 30-12-96 & 4$\times$ 90, 4$\times$180\\
              &             &&                 & 31-12-96 & 4$\times$ 90 \\
He 3-230      & PK 266-00.1 &  -    & Be       &30-12-96  & 24$\times$300 \\
HD 87643      & He 3-365    &  8.5  & B3/4[e]  & 31-12-96 & 20$\times$ 90\\
              &             &&                 & 01-01-97 & 12$\times$120 \\
MWC 297       &             &  12.3 & B1.5Ve   & 15-07-98  & 56$\times$100 \\
\hline
\hline
\end{tabular}
\ , \\
\noindent
Spectral types taken from Th\'e \ea\ (1994), except for MWC 297, taken
from Drew \ea\ (1997). {\it
V} magnitudes taken from {\sc simbad}. The integration times denote the
{\it total} on-source exposures.

\end{table*}

\section{Observations}

\subsection{Sample selection}

The target stars were selected from the catalogue of Th\'e, de Winter
\& Perez (1994) which lists all objects that had been at that time
proposed to be Herbig Ae/Be objects, and provides tables of other
emission type objects whose nature is not clear.  The list of targets
is provided in Table~\ref{targets}.  The targets were not selected
with foreknowledge of envelope asphericity, rather, they were chosen
because of their relative brightness, their position on the sky, and
their early (B-type) spectral types.

\begin{table*}
\caption{Results
\label{log} }
\begin{tabular}{llllllll}
\hline
\hline
Object    & Date     & \ha\  EW (\ang)  &  line/cont &   P$_{cont}$ (\%) & $\Theta_{cont}$ (\degree) &  Line-effect? \\
\hline
\hline
HD 37806  & Jan 95      & --17             &   4.9   & 0.29 (0.01)  & 121 (1) & $\surd$ \\
          & Dec 96      & --22             &   4.8   & 0.36 (0.01)  & 125 (1) & $\surd$ \\
$\omega$ Ori &          &  --3.5           &   1.6   & 0.30 (0.01)  & 55  (1) & - \\
V380 Ori  &             & --79             &   14    & 1.26 (0.01)  & 96  (1) & -  \\
MWC 137   &             & --550            &  83     & 6.11  (0.01) & 162 (1) & $\surd$ \\
HD 259431 &             & --63             &   11    & 1.06 (0.01)  & 102  (1)& $\surd$? \\
HD 53367  &             & --14             &   2.6   & 0.49 (0.01)  & 44  (1) & $\surd$ \\
HD 45677  & Jan 95      & --200            &  35     & 0.33 (0.01)  & 11 (1)  & $\surd$ \\
          & Dec 96      & --200            &  34     & 0.14 (0.02)  & 143 (3) & $\surd$ \\
HD 50138  & Jan 95      & --67             &  13     & 0.71 (0.01)  & 161 (1) & $\surd$ \\
          & Dec 96      & --58             &  13     & 0.65 (0.01)  & 154 (1) & $\surd$ \\
AS 116    &             &  --90            & 18      & 1.41 (0.01)  & 30  (1) & -  \\
LK \ha\ 218  &          & --20             &   6     & 1.91 (0.02)  & 19  (1) & -? \\
HD 52721  &             & --14             &   3     & 1.15  (0.01) & 19.0 (1)& - \\
HD 76534  & Jan 95      & --7.5            &   2.0   & 0.52 (0.01)  & 124 (1) & - \\
          & Dec 95      & --4.0            &   1.8   & 0.49 (0.03)  & 122 (2) & - \\
          & 30 Dec 96   & --6.1            &   1.9   & 0.50 (0.01)  & 125 (1) & - \\
          & 31 Dec 96   & --3.7            &   1.7   & 0.47 (0.02 ) & 128 (1) & - \\
Hen 3-230 &             & --315            & 64      & 1.61 (0.02)  & 30 (1)  & - \\
HD 87643  & Dec 96      & --186            &  25     & 0.84 (0.01)  & 168 (1) & $\surd$ \\
          & Jan 97      & --196            &  26     & 0.75 (0.01)  & 164 (1) & $\surd$ \\
MWC 297   &             &  --520           & 100     & 1.90 (0.01)  &  86 (1) & -\\
\hline
\hline
\end{tabular}
\ , \\
\phantom{tjeempie}
Continuum defined from 6400-6500 and 6600-6700 \ang .  The errors on
the equivalent widths of the \ha\ lines are expected to be of order
5\%, the errors cited for the polarization data are based on
photon-statistics only, and rounded upwards.  Systematic
(i.e. external) errors in the polarization are estimated to be of
order 0.05\% -- 0.1\%.
\end{table*}

\subsection{Spectropolarimetry}

The optical linear spectropolarimetric data were obtained using the
RGO Spectrograph with the 25cm camera on the 3.9-metre
Anglo-Australian telescope during three observing runs in January
1995, December 1995 and December 1996 respectively. During the first
two runs, the weather provided some spectacular views of lightning
from the telescope, but only limited data.  During clear time, we
aimed at observing the brightest objects in order to make the best of
lower-than-desired count rates.  Nevertheless, the resulting
polarization measurements proved to be very stable.  The last run was
mostly clear, opening the way for time to be spent on some of our
fainter targets.

The instrumental set-up was similar during all observing runs and
consisted of a rotating half-wave plate and a calcite block to
separate the light into perpendicularly polarized light waves.  Two
holes of size 2.7 arcsec and separated by 22.8 arcsec in the dekker
allow simultaneous observations of the object and the sky.  Four
spectra are recorded, the O and E rays of the target object and the
sky respectively.  One complete polarization observation consists of a
series of consecutive exposures at four rotator positions.  Per
object, several cycles of observation at the four rotator positions
were obtained in order to check on the repeatability of the results.
Indeed, we find that multiple observations of the same star result in
essentially the same polarization spectrum.  To prevent the CCD from
saturating on the peak of \ha , shorter integration times were adopted
for those objects with particularly strong \ha\
emission. Spectropolarimetric and zero-polarization standards were
observed every night.

A 1024 $\times$ 1024 pixel TEK-CCD detector was used which, combined
with the 1200V grating, yielded a spectral range of 400 \ang, centered
on \ha. Wavelength calibration was performed by observing a
copper-argon lamp before or after each object was observed.  In all
observations reported here a slit width of 1.5\arc\ was used.
A log of the observations is provided in Table 1. 
Bias-subtraction, flatfielding, extraction of the spectra and
wavelength calibration was performed in {\sc iraf} (Tody 1993).  The
resulting spectral resolution as measured from arc lines is 60 \km .
The E and O ray data were then extracted and imported into the Time
Series/Polarimetry Package ({\sc tsp}) incorporated in the {\sc
figaro} software package maintained by {\sc starlink}.  The Stokes
parameters were determined and subsequently extracted.

A slight drift of a few degrees in position angle (PA) was calibrated
by fitting its wavelength dependence in nightly 100\% polarized
observations of bright unpolarized stars (obtained by inserting an
HN-22 filter in the light-path) and removed from the polarization
spectra.  The instrumental polarization deduced from observations of
unpolarized standards proved to be smaller than 0.1\% in all cases.

In July 1998, MWC 297 was observed in service time with the ISIS
spectrograph and polarimetric optics on the 4.2m William Herschel
Telescope, La Palma. The instrumental set-up included the 1200R
grating and a 1124$\times$1124 TEK2 detector, providing a wavelength
coverage of 400 \ang \ around \ha \ and a spectral resolution of 40
\km . The data reduction was the same as for the AAT data.

Polarization accuracy is in principle only limited by
photon-statistics.  One roughly needs to detect 1 million photons per
resolution element to achieve an accuracy of 0.1\% in polarization
(the fractional error goes at $\sim$ 1/$\sqrt N$).  However, although
it is probably fair to say that the internal consistency of a
polarization spectrum follows photon-statistics, the external
consistency (i.e.  the absolute value for the polarization, checking
for variability) is limited due to systematic errors.  For example,
when calculating the polarization of a given spectral interval one can
reach polarization percentages with a statistical error of several
thousandth of a percent.  However, instrumental polarization (less
than 0.1\%), scattered light and low-level intrinsic variability of
the polarization standards may influence the zero-points.  The quality
and amount of data taken of spectropolarimetric standard stars is at
present not yet sufficient to reach absolute accuracies below the
0.1\% mark (see manual by Tinbergen \& Rutten, 1997).  A feeling for
the possible accuracies in our data can be obtained by studying some
of the objects that have been observed on different occasions. Seven,
respectively 6 independent observations of the Be star HD 76534 and
the polarization standard HD~80558 yield a mean polarization and
rotation of (0.49 \% with an r.m.s. scatter of 0.03 \%, 124\degree \
with a scatter of 3\degree) and (3.19 $\pm$ 0.11\% , 162 $\pm$
1.6\degree ) respectively.  It is encouraging to note that our
independent continuum measurements stretching over more than a year
are mostly within 0.1\%, and often within 0.05\%, in polarization.

\section{Results}

Some \ha \ parameters and continuum polarizations are presented in
Table~2.  In the following, the results across the full range of
targets observed are summarized.  These are grouped such that we begin
with those objects showing no discernable polarization changes across
\ha\ (\S 3.1), and then move on to objects which do show percentage
changes and/or rotations (\S 3.2).

Unless specifically stated, we have made no attempt below to correct
for the interstellar polarization (ISP). This decision is based on the
following: the main goal of this study is the detection of
polarimetric changes across the \ha\ line. Since the wavelength
dependence of the interstellar polarization  only becomes apparent on
wavelength ranges larger than our spectra provide, the ISP will only
contribute a constant polarization vector in (Q,U) space to the
observed spectra.  A further reason to refrain from ISP corrections
here, is that the methods commonly used for this (field-star method
and continuum variability, see e.g.  McLean \& Clarke 1979) do not
always return unambiguous values.  However, in the absence of ISP
correction, it is useful to remember the point raised in the
introduction that the ISP can change what might otherwise be a
reduction of the linear polarization percentage across the \ha\ line
into an increase in polarization, or an apparently constant
polarization, but accompanied by a significant rotation in the
position angle. The same effect can occur in the event of additional
polarization due to circumstellar dust.

Nevertheless, regardless of the influence of the ISP and polarization
due to circumstellar dust, it is possible to derive the {\it
intrinsic} angle of the electron-scattering material
(e.g. Schulte-Ladbeck \ea 1994). Assuming the line is depolarized, the
vector connecting the line- and continuum polarization in the QU plane
will have a slope that is equivalent to the intrinsic angle of the
scattering material responsible for the continuum polarization.  Since
the wavelength dependence of both circumstellar dust polarization and
ISP is small, they add only a constant QU vector to all points
in both line and continuum, and thus will not affect the
difference in line-to-continuum polarization.  This slope is measured
as $\Theta$ = 0.5$\times$atan($\Delta$U/$\Delta$Q).

\begin{figure*}
\mbox{
\epsfxsize=0.26\textwidth\epsfbox{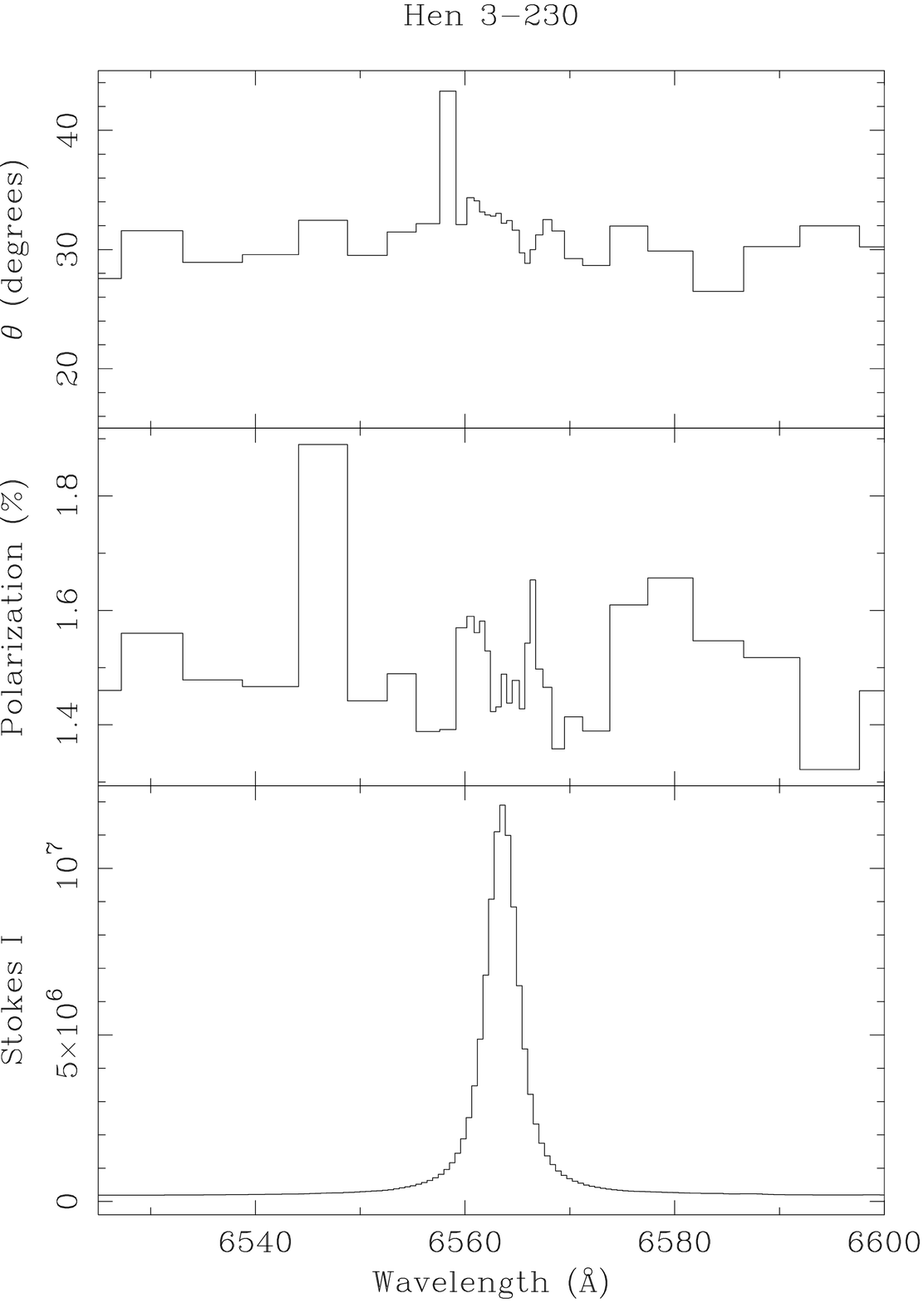}
\epsfxsize=0.26\textwidth\epsfbox{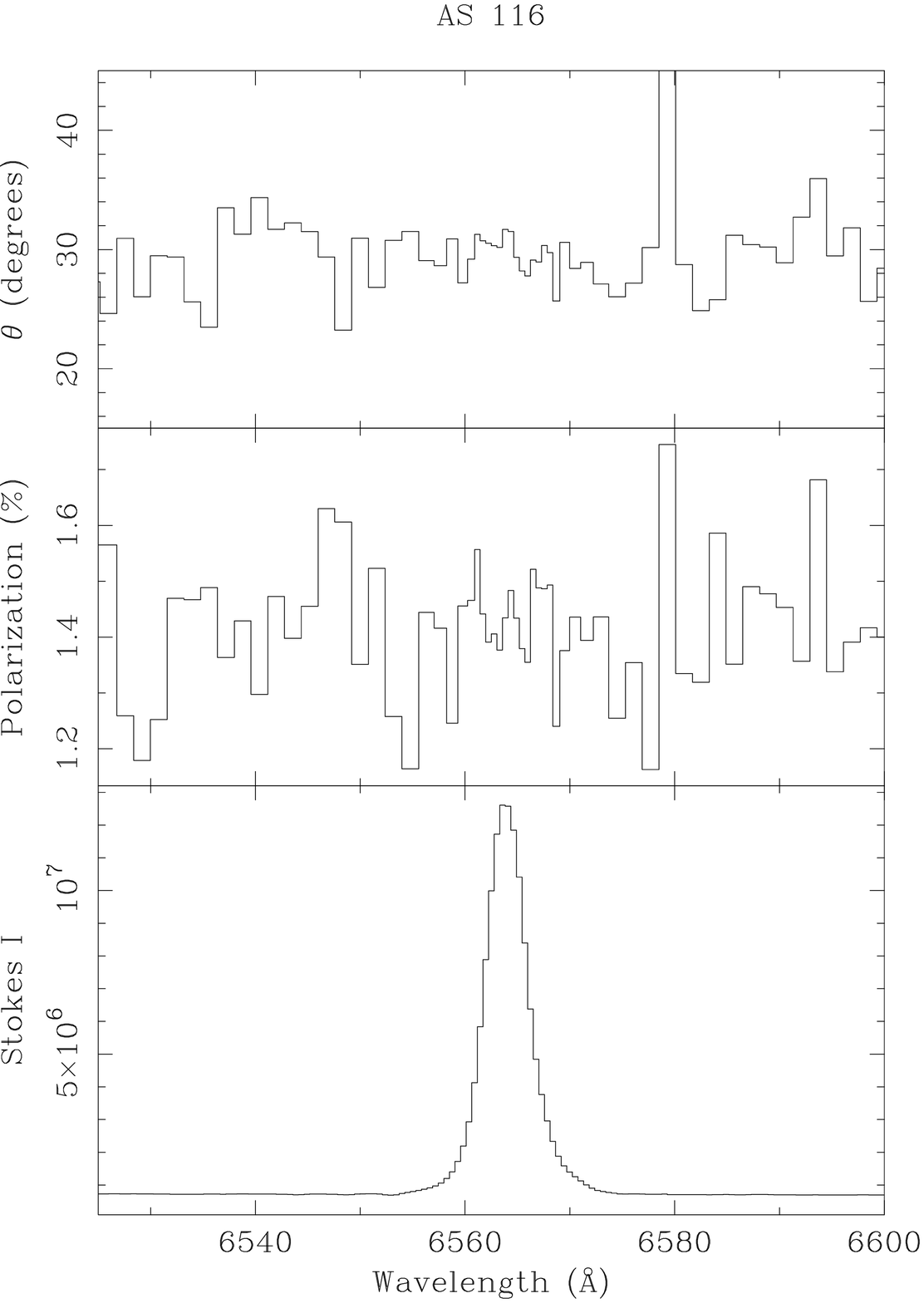}
\epsfxsize=0.26\textwidth\epsfbox{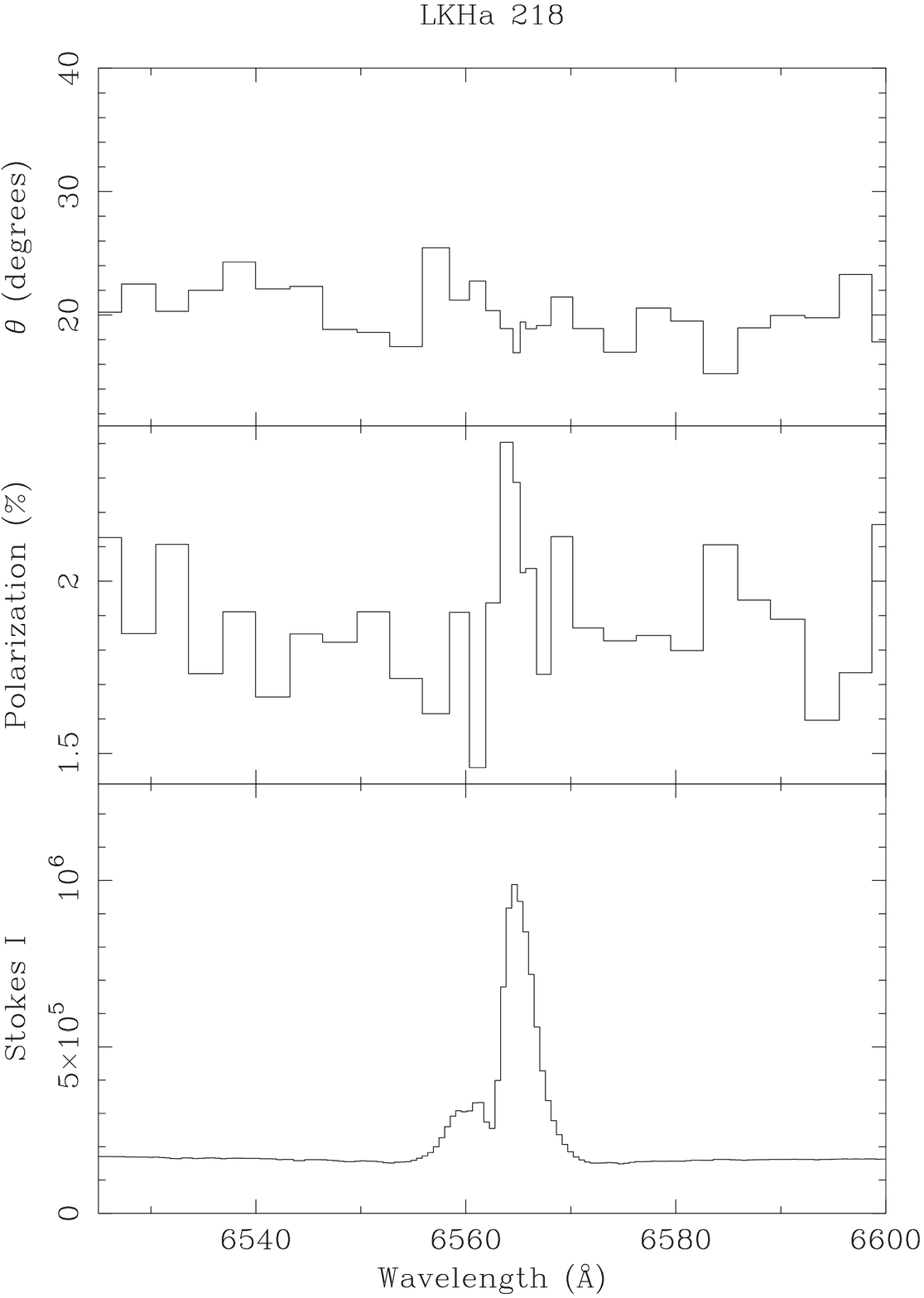}
\epsfxsize=0.26\textwidth\epsfbox{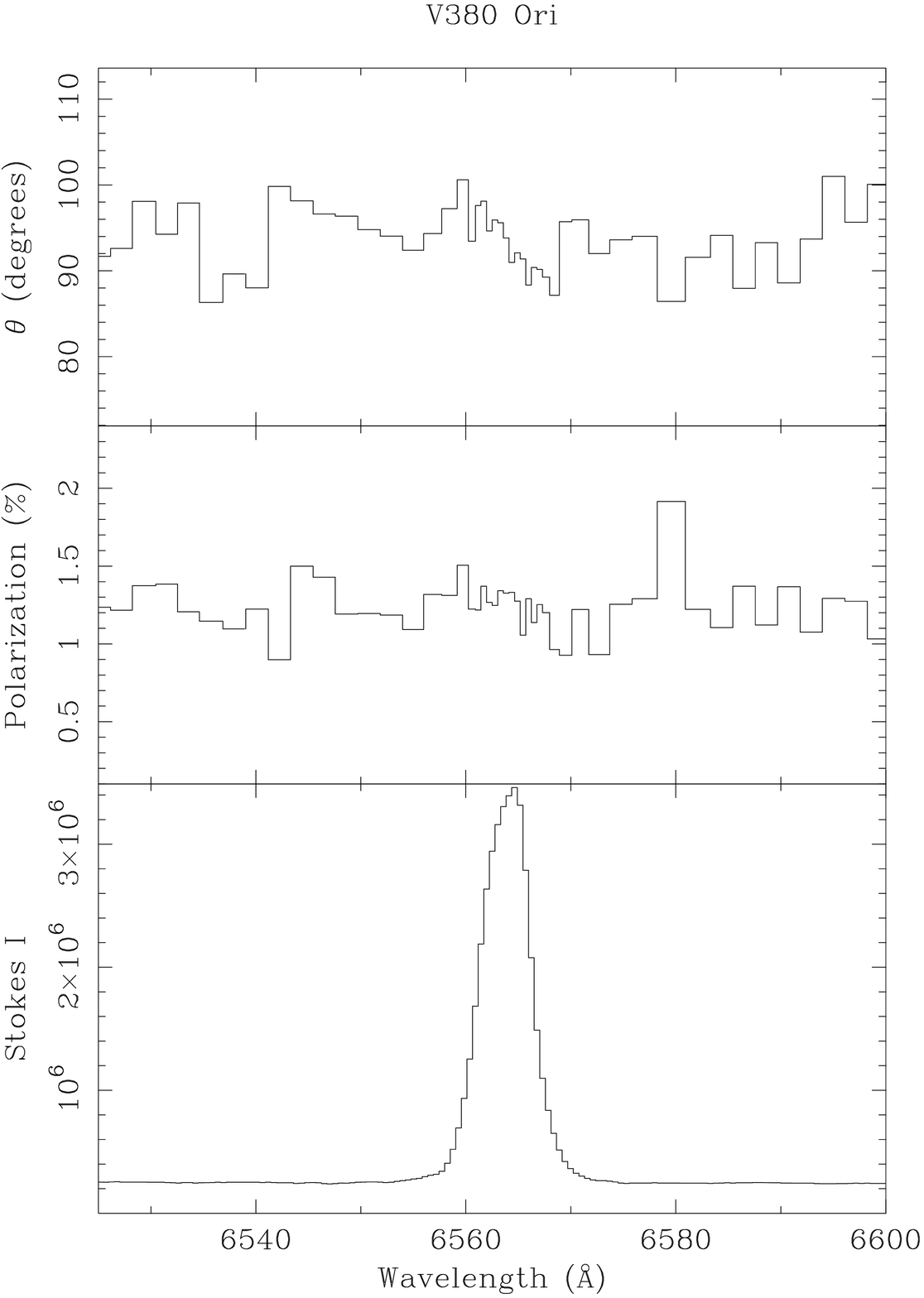}
}
\mbox{
\epsfxsize=0.26\textwidth\epsfbox{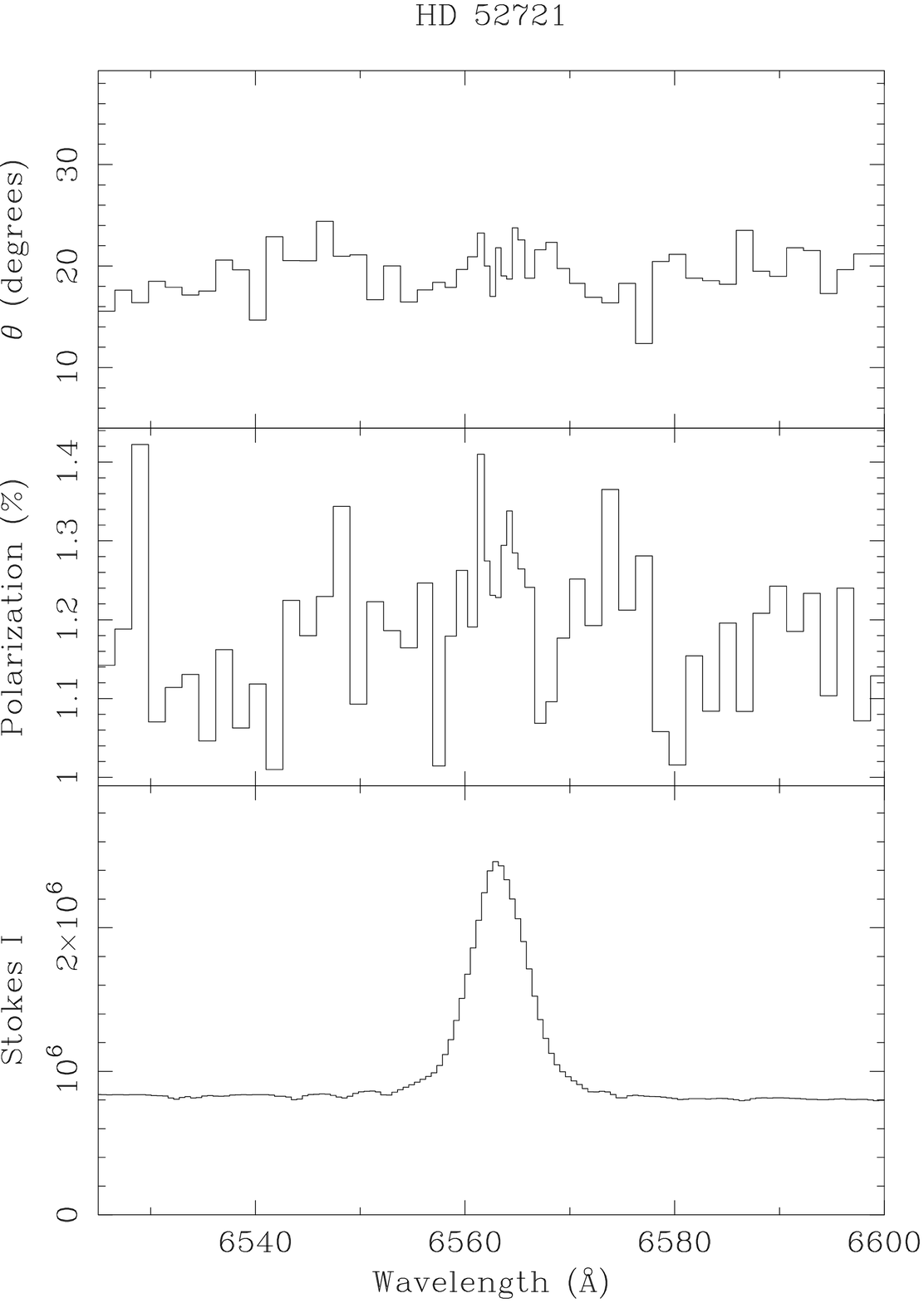}
\epsfxsize=0.26\textwidth\epsfbox{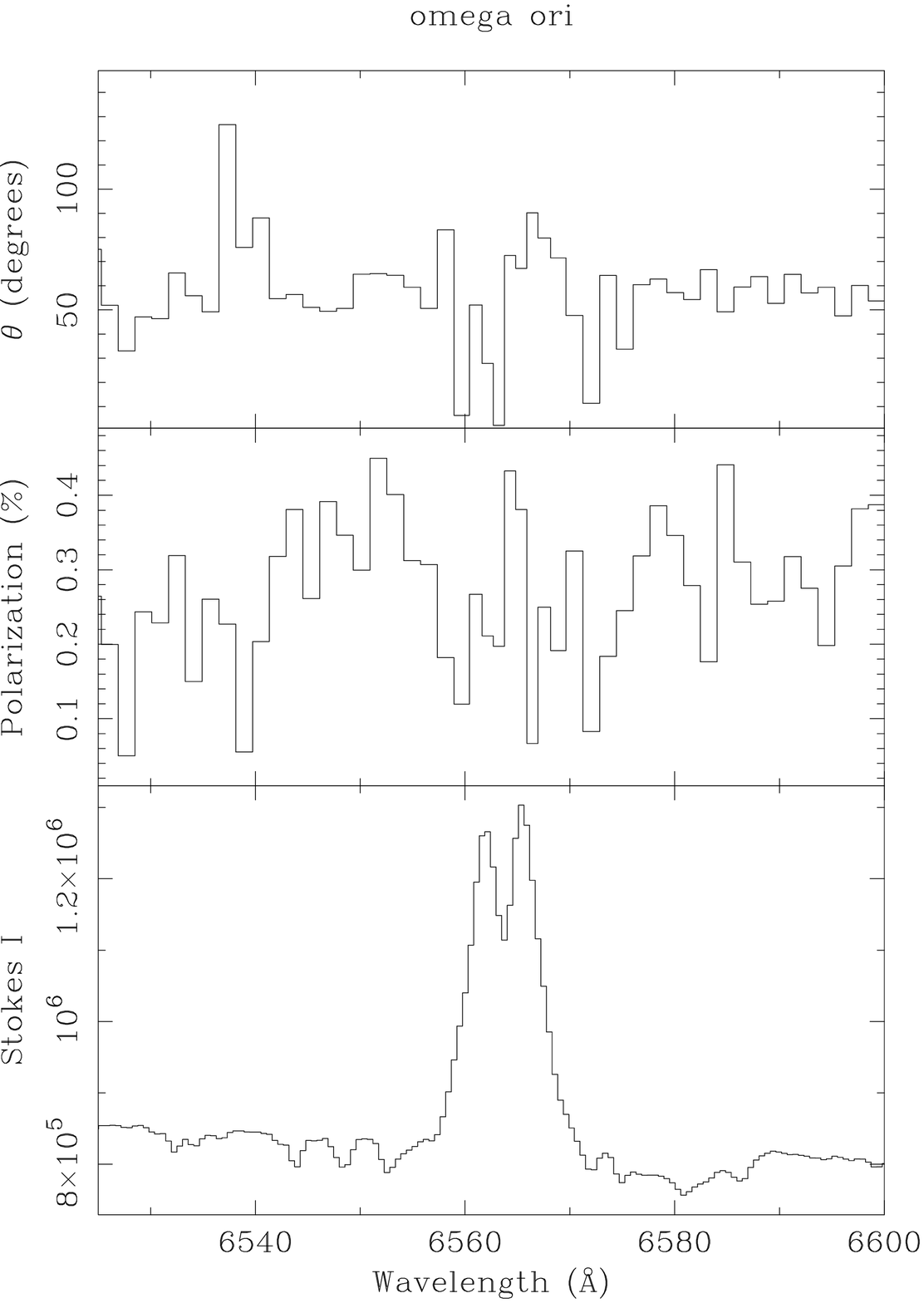}
\epsfxsize=0.26\textwidth\epsfbox{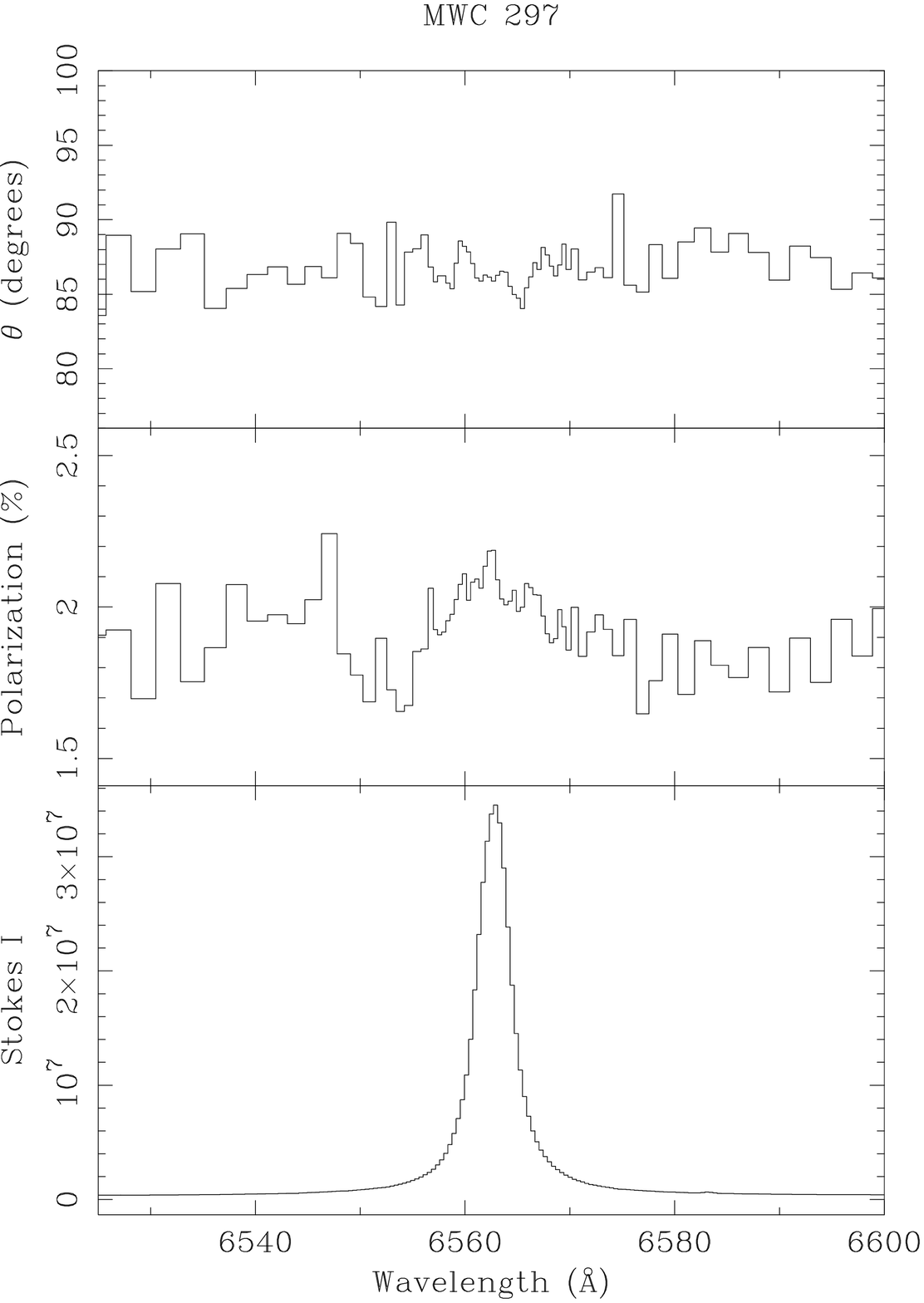}
\epsfxsize=0.26\textwidth\epsfbox{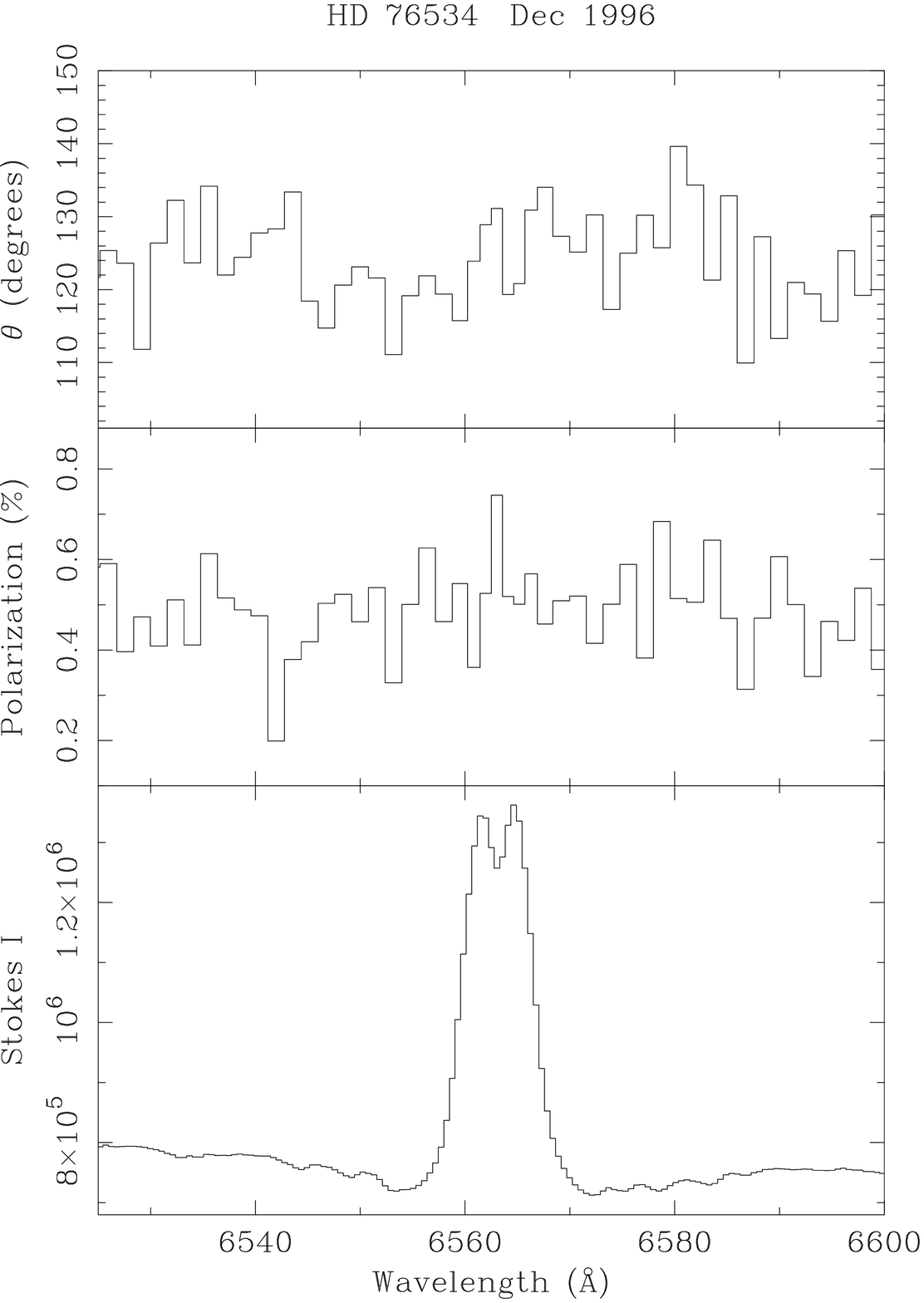}
}
\caption{
Polarization spectra of the stars showing no clear polarization 
change across \ha.
In each case, the normal (intensity) spectrum is shown in the lower
panel, the polarization (in \%) in the middle panel, while the
position angle is plotted in the upper panel. In each case and 
in the following figures (unless stated otherwise)   the data
are  rebinned such that the 1$\sigma$ error in the polarization
corresponds to 0.1\% as calculated from photon statistics.
\label{nondet}
}
\end{figure*}

\subsection{Stars showing no clear change across \ha }

In this subsection we discuss the objects that do not show a
line-effect.  In principle such an observation implies that the
projection of the ionized region on the plane of the sky is (mostly)
circular. We will find that this does not necessarily have to be the
case.  The objects falling into this group are Lk~\ha ~218, Hen~3-230,
AS~116, HD~52721, V380~Ori, HD~76534, $\omega$~Ori and MWC~297.  Their
polarization spectra are shown in Figure ~1.

\paragraph*{Hen~3-230, AS~116, Lk~\ha ~218}
These objects are relatively faint targets for which the
signal-to-noise ratios in our data are not so high.  Hence, the
absence of change across \ha\ for the time being should be viewed as
an absence of any marked contrast.  For example in the case of Lk~\ha
~218, there seems enhanced polarization at the position of the
stronger redshifted emission component in the line.  However this is
not strictly even a 2$\sigma$ detection.  Coarser binning can yield a
3$\sigma$ enhanced polarization in the line (at  smaller resolution
this is only present in one pixel however) but in truth it would appear that
100 minutes exposure is not enough for this object.  The null results
for Hen~3-230 and AS~116 are more sure.  Both targets have extremely
bright line emission.  As previous observations of them are extremely
sparse, their evolutionary status remains undetermined.  Based on its
low excitation spectrum Stenholm \& Acker (1987) argue that Hen 3-230
is not a Planetary Nebula, despite having figured in many previous
papers to be one.  AS 116 was appeared in a catalogue of emission line
stars of Miller \& Merrill (1951), since then not much work has been
published.  The IRAS flux peaks at 25\mic, which could point at a
detached dust shell, but it was not detected in the OH maser by
Blommaert, van der Veen \& Habing (1993).

\paragraph*{HD~52721, V380~Ori}
These are two quite convincing examples of no line effect.
Both, nevertheless, present significant continuum polarizations.  Since
HD~52721 presents little of an infared continuum excess (Hillenbrand et
al. 1992), it might seem plausible that this star is a low-inclination 
classical Be star behind a significant interstellar column.  Indeed the
single-peaked H$\alpha$ emission is consistent with this, but   
in contrast to the $v\sin i$ measurement of 400$\pm$40 km s$^{-1}$
reported by Finkenzeller (1985) 
suggesting high inclination.  V380~Ori
exhibiting a strong infrared continuum excess, would appear to be
optically-veiled and hence sits more convincingly in the HAeBe object
class.  The absence of a line-effect in V380~Ori may simply imply lower
inclination to the line of sight.

\paragraph*{HD 76534}
    The initial 1995 data on this source have already been presented
by Oudmaijer \& Drew (1997).  There it was shown that these data
did not indicate any changes across \ha . Our new data confirm this
and show no hints of pronounced polarization variability in its modest
$\sim$0.5\% level.  This is despite the source's propensity for
spectral variability clearly illustrated by the 11 January 1995
transformation of \ha\ absorption into well-developed double-peaked
emission within hours (Oudmaijer \& Drew 1997).  Table~2 lists the \ha
\ equivalent widths at the various occasions that the object was
observed. The EW changes strongly, but situations similar to the
January 1995 data were not observed.

\paragraph*{$\omega$~Ori}
    It is not clear whether $\omega$~Ori should be considered a Herbig
Be star, or simply a classical Be star (Sonneborn \ea\ 1988).  The
absence of a detectable change across \ha\ (Fig.~\ref{nondet}) stands
in contrast to reports in the literature that the hydrogen
recombination lines show de-polarization -- PM find that \ha\ shows a
polarization dip while Clarke \& Brooks (1984) find the same in
H$\beta$.  This difference is presumably connected with the stronger
\ha\ emission reported by PM (line to continuum ratio of 1.8 versus
our figure of 1.4 - which was not binned to the same narrow band, and
is thus a strong upper limit), and higher linear polarization (0.38\%
versus our 0.30\%).
   Since classical Be stars and, indeed, Herbig Be stars are known to be
emission line variables, this change is probably due to a lowering of
the ionized emission measure of the equatorial disk around this star.
Given the great disparity between the Thomson scattering and H$\alpha$ 
absorption cross-sections, a relatively modest drop in the H$\alpha$ 
equivalent width could well be accompanied by a collapse in the percentage 
of linearly-polarized scattered starlight.  Hence it would seem that
ISP contributes around 0.3\% linear polarization in $\omega$~Ori, a figure
not out of line with PM's estimate of 0.24\% .

\paragraph*{MWC~297}

    The weak-to-non-existent effect across \ha\ is startling in view
of the evidence gathered by Drew et al. (1997) that this early Herbig
Be star is viewed at relatively high inclination.  Furthermore, the
5 GHz radio image (see Drew et al. 1997), which provides an extinction-free
view of the ionized circumstellar medium around MWC~297, indicates an
elongated geometry that would suggest a line effect ought to be
apparent in such a bright emission line source.

Although at first sight very surprising, we consider two different
hypotheses that may explain this apparent paradox.
Firstly, we can conclude we are seeing the \ha \ line directly, and
that the line-forming region is indeed round.  Since the \ha \ line is
formed in a potentially much smaller volume than the continuum 5 GHz
radiation, the rounder appearance of the \ha \ line-forming region
indicates that the geometry changes between the near-stellar scale and
the larger scale sampled at radio wavelengths.  Spatial evolution of
this type has been predicted for lower mass stars  (see Frank
\& Mellema 1996).

Secondly, it may be that \ha \ is formed in an edge-on disk like
structure, but that the optical light does not reach us directly, and
is completely obscured in the line-of-sight.  The light that we see
could then be `mirrored' by scattering dust clouds located above
and/or beneath the obscuring material.  If the scattering dust-clouds
`see' a nearly circularly symmetric \ha \ emitting region, it will not
see any de-polarization across the line either.  Consequently, the
light reaching us will not show any polarization changes across \ha .
That dust-scattering plays a role in this object is already suggested
by the spectral energy distribution, which shows a notable excess in
the $U$-band (Bergner \ea \ 1988; Hillenbrand et al 1992), possibly
due to the `blueing' effect.

We may therefore have a similar situation to that in the Red Rectangle
(see e.g. Osterbart, Langer \& Weigelt 1997, Waelkens et al 1996),
where it was only recently realized that the central star is actually
not the star itself, but its reflection against dusty knots located
above and below a very optically thick dust lane.  This finding
explained the long-standing problem of the energy balance; the
apparently absorbed light from a star with such a modest reddening
($A_V$ of order 1) is orders of magnitude less than that being
re-radiated in the infrared.

If the circumstances are similar in MWC 297, the reddening (\Av \
$\approx$ 8 - see discussion of Drew et al, 1997) commonly assigned to
this source on the basis of conventional extinction measurements is a
severe underestimate.  This would not be completely unexpected, as MWC
297 is in certain respects an intermediate object between the
optically visible Herbig Be stars, and their more massive
counterparts, the Becklin-Neugebauer (BN) type objects, which suffer
from large optical extinctions (\Av \ often in excess of 20). While in
the optical, MWC 297 has much in common with the Herbig Be stars, at
infrared and radio wavelengths it shows evidence of substantial mass
loss associated with BN-type objects (Drew \ea\ 1997).

\subsection{Objects displaying line effects} 

Here we present the objects for which the line-effect is
observed. First, the Herbig Be stars in this sample are discussed, HD
259431, HD 53367 and HD 37806, then MWC 137, a Herbig Be star that has
been recently proposed to be a massive evolved B[e] star instead is
shown, and we end with the well-known B[e] objects HD 50138, HD 87643
and HD 45677.

\paragraph*{HD 259431}

\begin{figure}
\begin{center}
\mbox{
\epsfxsize=0.26\textwidth\epsfbox{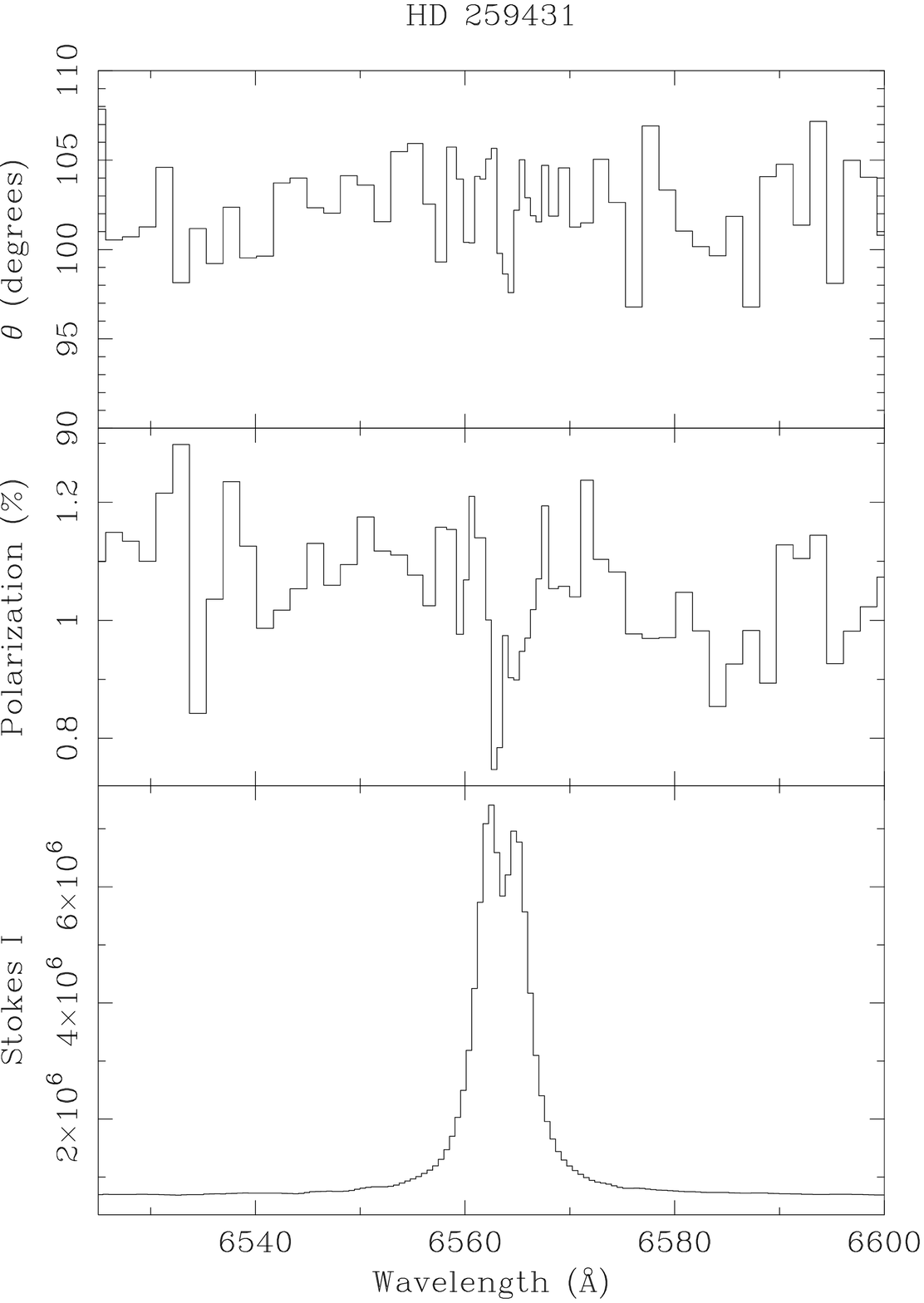}
\epsfxsize=0.26\textwidth\epsfbox{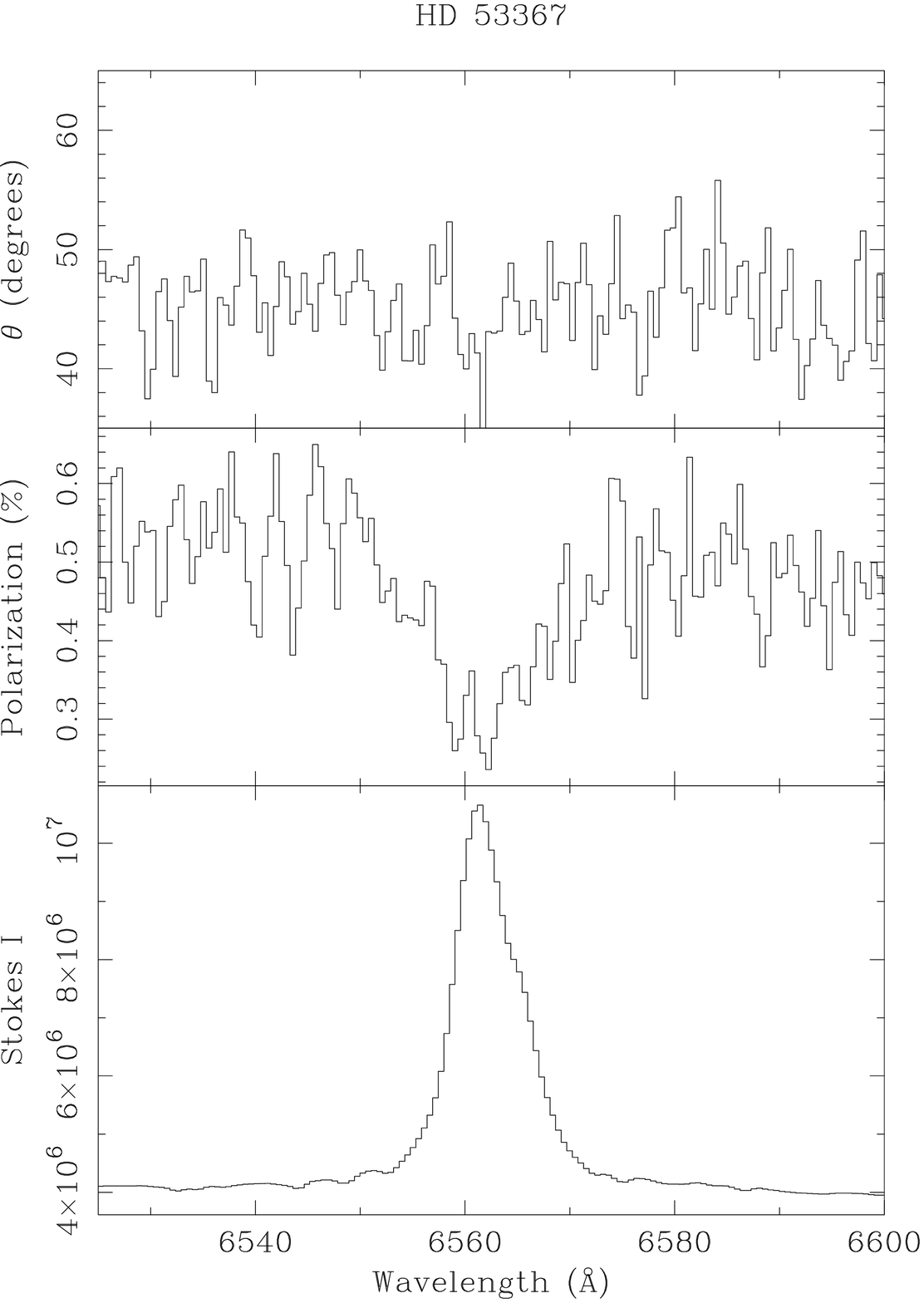}
}
\end{center}
\caption{
The polarization spectra of HD 259543 and HD 53367.
\label{259533}
}
\end{figure}

We start with the least certain detection, HD 259431.  This object
(Fig.~\ref{259533}) shows a hint of de-polarization across \ha , from
(1.1\%, 102\degree) to 0.8\% in the line center.  The intrinsic
polarization angle in QU space measured from the change from the
continuum to line polarization, $\Theta$ =
0.5$\times$atan($\Delta$U/$\Delta$Q), gives 17\degree, but with a large
uncertainty.  The length of this vector is small, of order 0.3\%.  Our
two observations, taken one year apart, do not show any variability
within the small error-bars. The compilation by Jain \& Bhatt (1996)
which contains broad-band polarimetric observations only hints at a
slight variability. The high resolution data, only when binned to 
errors of 0.1\% or less, show the line-effect.

\paragraph*{HD 53367}

The polarization spectrum of HD 53367 is shown in Fig.~\ref{259533}.
Since there was no difference within the errorbars of the data taken
on 2 consecutive nights, these were added to increase the
signal-to-noise ratio.  The \ha\ line is clearly de-polarized with
respect to the continuum, while no rotation across the line is
present.  The intrinsic polarization angle in QU space measured from
the slope is 47\degree .

If the line center is assumed to be completely de-polarized, one can
use this information to correct the observed polarization for the ISP
(the `emission line method').  Reading off the polarization in the
line-center (Q = 0.0\%, U = 0.3\%), and subtracting this value from
the spectrum then gives an intrinsic polarization of 0.2 $\pm$ 0.01 \%
and a position angle of 44.5 $\pm$ 0.5\degree \ (measured in the bins
6400 -- 6500 \ang\ and 6700 -- 6800 \ang . Note that the error-bar
reflects the internal consistency and not the external consistency),
consistent with the slope in the QU plane.  This low value of
intrinsic continuum polarization is what one would expect from modest
electron-scattering (see e.g. PM).

Let us now comment on the significance of the measured polarization
angle.  On the sky, HD~53367 is located on the periphery of the Canis
Majoris `hole' noted for its low reddening sightlines (Welsh 1991).
Herbst, Racine \& Warner (1978) designated this star a member of the
CMa~R1 cluster, which they placed at a distance of 1150~pc, the same
distance as to the CMa OB1 association (Claria 1974).  This distance
is suspect as it makes HD~53367 more luminous than a supergiant at the
same B0 spectral type.  A more reasonable distance estimate would be
around 550~pc (adopting a dereddened $V$ magnitude of $\sim 4$;
Herbst et al. 1982, and $M_V \simeq -4.7$; for a B0IV star,
Schmidt-Kaler 1982).  It is then less surprising that the Hipparcos
catalogue (ESA, 1997) contains a finite, although very uncertain,
parallax measurement for this star (see also van den Ancker, de Winter
\& Tjin A Djie 1998).  In any event there is strong evidence that the
cumulative interstellar extinction towards CMa~R1 is not more than
$A_V \simeq 0.2$, implying that  the remaining observed extinction is local
(Herbst et al. 1982; Vrba, Baierlein \& Herbst 1987).  

More important,
Vrba et al. (1987) demonstrate quite convincingly that within CMa~R1
the polarization angle tends to follow the sweep of the southern dust
arc up through Sh2~292, the H{\sc ii} region ionized by HD~53367, and
that the polarization is most likely due to grain alignment.  At
HD~53367 this angle is about 40\degree\ and entirely comparable to
that of the non-emission line B3V star BD~-10$^o$1839 just
$\sim$20~arcmin away, and also at a photometric distance of about
600~pc.
The interesting feature of HD~53367 is that the separable intrinsic
and foreground polarization angles are the same -- as indicated by the
lack of rotation across H$\alpha$ in the observed spectrum.  This
suggests an orderly star formation process in which the rotation axis
of HD~53367 `remembers' the larger scale circumstellar field
direction, in preference to a more dynamical mechanism such as the
`accretion induced collision' merger model of Bonnell et al (1998),
which would result in randomly oriented polarization angles.

\paragraph*{HD 37806}

\begin{figure}
\mbox{\epsfxsize=0.52\textwidth\epsfbox{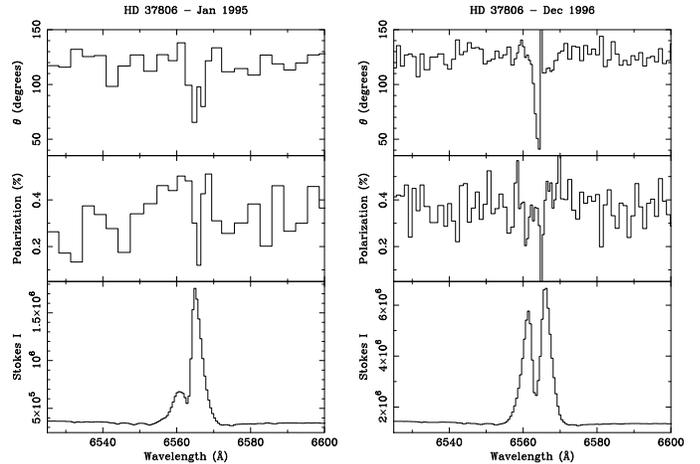}}
\caption{Observed spectra of HD 37806 in 1995 (left hand panel) and
1996 (right hand panel). 
\label{h378}
}
\end{figure}

The double peaked \ha\ profile of this object is remarkably different
during our two observing epochs (Fig.~\ref{h378}).  In December 1996,
both peaks were equally bright, while in January 1995 the blue peak is
much weaker.  Although the signal-to-noise of the January 1995 data is
not high, it appears that the object does not show significant changes
in polarization.  But a rotation is present, which is especially
visible in the December 1996 data when the source was observed for
longer.  The rotation almost exactly occurs in the central dip in \ha
, rather than across the entire adequately-resolved line profile.  The
fact that we observe only rotation is interesting in its own right, as
the principle of line de-polarization without superimposed foreground
polarization would imply a constant angle across \ha .  Clearly there
is ISP and perhaps circumstellar polarization present.

\begin{figure}
\mbox{\epsfxsize=0.52\textwidth\epsfbox{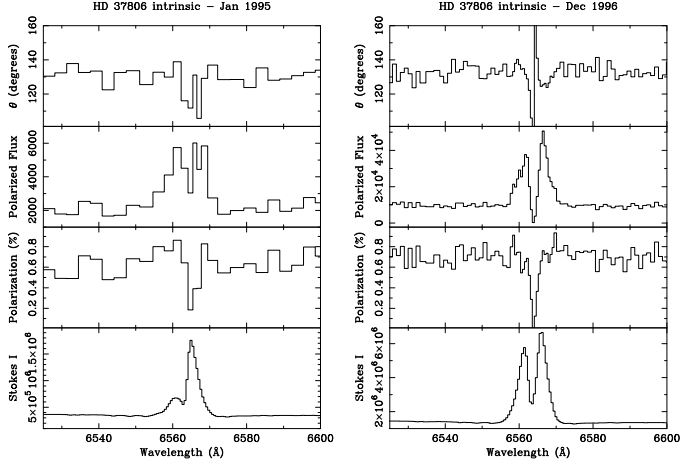}}
\caption{
Intrinsic spectra of HD 37806 obtained  by assuming the line-center to be
intrinsically unpolarized. The second panel shows also the polarized flux
(polarization $\times$ intensity).
\label{h378int}
}
\end{figure}

If we assume that the underlying, rotated part, of the line-profile is
unpolarized, we may attempt a correction for the intervening
interstellar- and circumstellar dust polarization, to retrieve the
intrinsic spectrum of this object.  The (Q,U) vector measured in the
central dip of the rotation, corresponds to (-0.06\%,0.35\%) or a
polarization of 0.36\% with PA 50\degree. Subtracting these values
from the observed spectra results in Fig.\ref{h378int}. This figure
also shows the `polarized flux' (polarization $\times$ 
intensity). The polarized flux indicates that the double-peaked \ha\ line
has the same polarization as  the continuum, but -- by
virtue of the manner in which the ISP was corrected for -- the central
dip between the peaks is de-polarized.  The 1995 spectrum shows the
same behaviour. Despite its much lower signal-to-noise, the large
Red/Blue ratio has virtually disappeared in the polarized flux
spectrum, suggesting a large part of the red peak is  not
associated with the line-forming region responsible for the double
peaks.

A consistency check can be made as to whether the choice of ISP for
the correction is reasonable. We have searched the Matthewson \ea \
(1978) catalogue for field stars nearby the object and found 49
objects within a radius of 180 arcmin (one object with P = 12\% was
excluded for the analysis). The catalogue also provides
photometric estimates of the total extinction \Av \ to these objects.
A relatively tight relation exists between the observed polarization
and \Av .  A least-squares fit to the data gives the relation P (\%) =
1.5 $\times$ \Av (mag.)  + 0.07.  The PA shows mostly a scatter
diagram, and gives a mean of 76\degree $\pm$ 39\degree \ for the total
sample.  The \Av \ towards HD 37806 is ambiguous, but likely to be
low: Van den Ancker et al. (1998) reclassify HD~37806 as an A2Vpe star
and give its extinction as $A_V = 0.03$.  In contrast, Malfait,
Bogaert and Waelkens (1998) find an \ebv \ of 0.14 (\Av = 0.43 if the
ratio of total to selective reddening, $R$, is 3.1) for a B9 spectral
type.  Based on the extinction and the field stars, the ISP towards HD
37806 should be between 0.1\% and 0.7\%. The `emission line' method
gives 0.36\%, a value that is at least consistent with the value
returned from the field stars.

Taking the derived intrinsic polarization spectrum at face value, it
appears that the \ha \ line profile is a composite of two unrelated
components: a double-peaked polarized component and a single,
unpolarized component.  Since both the photospheric continuum
radiation and the double-peaked component are equally polarized, it
would appear that they both appear point-like to the scattering
material, while the single component is formed further away from the
star, betraying the asymmetry of the scattering region.  Perhaps due
to the signal-to-noise in the data, no de-polarization is visible in
the blue peak. This could suggest that the de-polarization in the red
peak is not necessarily due to electron scattering, since one would
expect the electron scatterers to be located close to the
star. Instead, the data do not exclude the possibility that the red
peak is located in an extended nebula (which is not resolved in our
data however), while the underlying broader emission and the
photosphere are polarized by circumstellar dust, which, by
implication, is not distributed spherically symmetric around the star.

Intriguingly, the line-to-continuum ratios of the red peak are
constant (see Table 2), while broad-band photometry of this object
also appears constant (Van den Ancker \ea \ 1998) so the blue peak has
increased in strength. This fact, combined with the relatively equal
red/blue ratio of the double-peaked line in polarized flux, suggests
that the polarized red part of the line has also become stronger.
This leads to the enigmatic situation that the polarized part of the
red peak increased in strength while the unpolarized part of the red
peak decreased in strength in such a way that their
total has remained constant in time.

Clearly, this object needs further study, both
spectropolarimetric, and from a modelling perspective to gain more
understanding as to the origin of  the observed polarization.

\paragraph*{MWC 137}

Although Th\'e et al. (1994) labelled this source as a probable Herbig
Be star, in a recent study  Esteban \& Fern\'andez (1998)  argued
that is much more likely to be an evolved B[e] supergiant.  Their
arguments, based on a kinematical association with molecular clouds at
more than 5 kpc, are reasonably convincing, but we note that the nature of
this source has long been controversial (see references in Esteban \&
Fern\'andez).

\begin{figure}
\mbox{\epsfxsize=0.50\textwidth\epsfbox{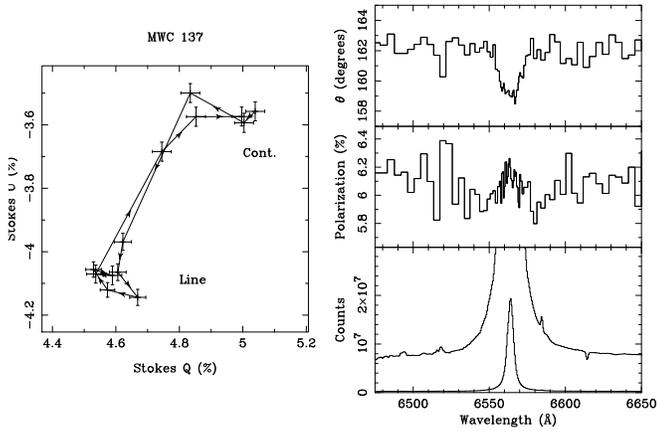}}
\caption{
The data on MWC 137. The left panel shows the QU diagram at 0.03\%
binning, a clear excursion from the continuum is present over the line
profile.  The right-hand panel shows the polarization spectrum.
The intensity is blown up by a factor of 30 for clarity, while
the wavelength range is expanded to show the broad line better.
\label{m137}
}
\end{figure}

The polarization spectrum and QU diagram of MWC 137 are shown in
Fig.~\ref{m137}. The continuum polarization of the object is very
large (6\%), and slight polarization changes across \ha\ are
visible. Much clearer is the broad, observed rotation of \ha ,
centered on the line peak.  The rotation is at most only 3\degree ,
but it is real as is evident from the significant loop apparent
between line and continuum in QU space (Fig.~\ref{m137}); the shift is
along the QU vector (+0.3\%, +0.5\%), corresponding to an intrinsic
angle in the continuum of 30\degree \ and a depolarization of
$\approx$ 0.6\% (measured from the length of the polarization vector,
$\sqrt{\Delta Q^2 + \Delta U^2}$).  The same situation as for HD 37806
described above occurs in this case: the polarization from intervening
material has transformed a de-polarized line into a rotated line.
Slight changes in the observed polarization are still present in the
wings of the emission, suggesting a different polarization of the line
wings than both the continuum or the centre of the emission line.

The interstellar reddening towards the object is very large. It was
redetermined by Esteban \& Fern\'andez (1998) to be \Av\ = 3.77.
This is very high, but according to the authors consistent with a very
large distance to the object (the authors mention 6 kpc).  If true,
then the large continuum polarization can be  explained mostly by the
interstellar reddening. The polarization of the field stars within a
radius of 300\arcm \ (taken from Matthewson \ea\ 1978) increases
linearly with \Av \ up to 4\% at \Av \ $\sim$ 2.2, the largest \Av \
among the field stars in the sample.  Unfortunately no information for
stars more distant or more reddened is available, but it is clear that
a large ISP can be expected for the object.

The rotation across the \ha \ line is best interpreted as a
depolarization across the line, but modified by the intervening ISP
and circumstellar dust polarization. Using the `emission line' method,
we find a dust polarization of P=6.2\%, $\Theta$=159\degree , in
agreement with the large expected ISP.  The intrinsic PA of the
electron scattering medium of 30\degree \ appears to be parallel with
the bright North-Western component of the ring nebula around the
object (see again Esteban \& Fern\'andez 1998). This may suggest that
asymmetries at very small scales, traced by the electron scattering,
and at large scales (from the image) are still aligned.

\paragraph*{HD 50138}

In their search for Herbig Ae/Be stars, Th\'e \ea\ (1994) found a
subsample of objects with many Herbig characteristics, that
nevertheless did not fulfil all their criteria. Based on the strong
emission lines, they called this group `extreme emission line
objects'.  Most of these stars are also classified as B[e] stars,
because of the presence of forbidden lines in the spectrum.  HD~50138
is one of these.

\begin{figure}
\mbox{\epsfxsize=0.48\textwidth\epsfbox{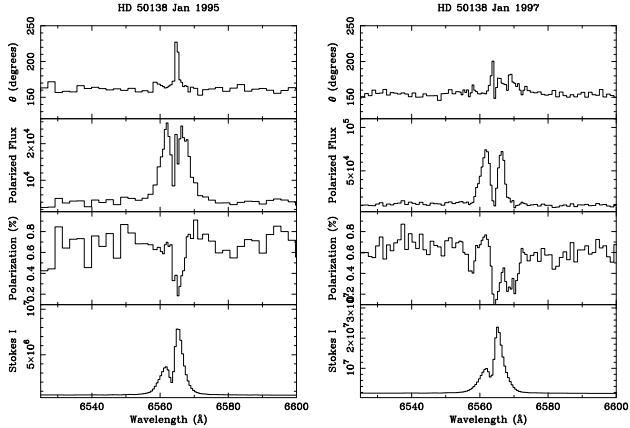}}
\caption{
The polarization data on HD 50138 in 1995 and 1997 respectively, the polarized flux is also plotted.
\label{h501pf}
}
\end{figure}

Our two observations of HD 50138 (Fig.~\ref{h501pf}), taken two years
apart in January 1995 and January 1997, show essentially the same
behaviour. The emission line is double peaked, with a large red to
blue ratio of the peaks.  The velocity separation of the two peaks is
160 \km\, and the central minimum is at 18 \km\ (helio-centric), which
is 15 \km\ blueshifted from the forbidden [O{ \sc i}] line at 6363
\ang , for which we measure a central velocity of 33 $\pm$ 5 \km
(helio-centric), in agreement with the radial velocity determination
by Pogodin (1997).

The \ha\ line shows strong de-polarization across the red
peak, while the blue peak shows at most a slight de-polarization. 
The `intrinsic' polarization angle, as deduced from the
shift between line and continuum in QU space is about 155\degree,
close to the measured one, implying that the ISP, if any,  has had  no
great effect on the observed polarization characteristics.  This is
more or less in keeping with the moderate reddening towards this
source (van den Ancker et al.  1998 assign $A_V = 0.59$, possibly an
upper-limiting value considering that $B-V$ for this mid-B star is
close to zero) and the  low interstellar polarization in the
line of sight, as many objects around HD 50138 have very low
polarizations for typical extinction values as 0.5 (Matthewson \ea \ 1978).

The polarized flux spectrum (polarization $\times$ flux,
Fig.~\ref{h501pf}) reveals equally strong blue and red peaks at both
epochs.  This may indicate that part of the red emission is formed in
the same region as the blue peak, such as a rotating disk, but that
the excess emission compared to the blue line is formed in a larger
volume, resulting in the observed de-polarization. The most
straightforward explanation to account for the `excess' flux in the
red peak, as in the case of HD 37806, is that the intensity spectrum is a
composite of a rotating disk type of geometry close to the star, and
an additional, extended single peaked component.  A clue to the line
formation may be provided by the spectrum taken around the lower
opacity H$\beta$ line by Jaschek \& Andrillat (1998).  Their published
H$\beta$ profile shows the blue and red peaks roughly equal, with some
underlying photospheric absorption still visible.  The much larger
red-to-blue ratio in \ha\ could imply that the excess red emission in
\ha\ is optically thin - since for thin H{\sc i} emitting gas the
Balmer decrement may be substantially steeper than for optically-thick
gas. The double-peaked part of the line could then be an
optically-thick line formed very close to the star, while the
unpolarized line is optically-thin.

It is clear that any picture of this object must be simplified since
the spectrum of HD 50138 shows many more peculiarities.  Grady et
al. (1996) include this star in their sample of objects exhibiting
$\beta$~Pic type infall phenomena.  Pogodin (1997) drew attention to
the variable He{\sc i} $\lambda$5876 line profile, which sometimes
shows an inverse P~Cygni behaviour.  He also attributed this to infall
of material.

\paragraph*{HD 87643 }

\begin{figure}\mbox{\epsfxsize=0.48\textwidth\epsfbox{
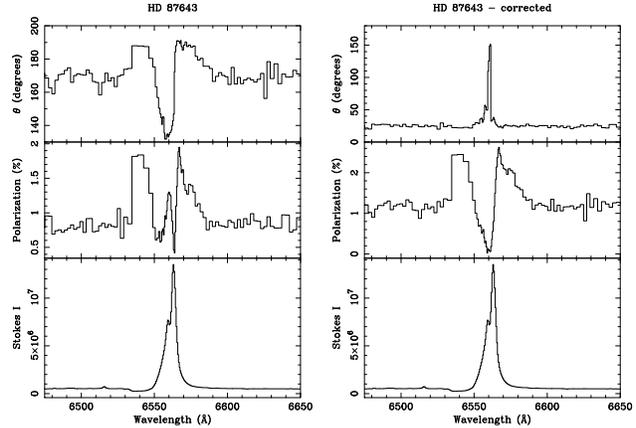}}
\caption{
The polarization spectrum of HD 87643, and the corrected spectrum of the star.
The wavelength range is expanded to show the broad line better.
For more details, see Oudmaijer \ea \ (1998).
\label{h87}
}
\end{figure}

HD 87643 is a B[e] star, for which evidence suggests that it is
located at several kpc, indicating a massive and evolved nature of the
object. This, and its spectroscopic and spectropolarimetric data have
been analysed in more detail in Oudmaijer \ea \ (1998). For
completeness we show the data and the spectrum corrected for ISP and
circumstellar polarization in Fig.~\ref{h87}.  The polarization of
this object shows some striking features. After correction for the
intervening polarization, it turns out that most of this structure
appears an artefact due to the polarization vector additions; the
spectrum corrected for ISP and circumstellar dust has a much smoother
behaviour. A comparison with the schematic model calculations by Wood,
Brown \& Fox (1993) indicates that the polarization profile can be
best reproduced with a circumstellar disk that is both rotating and
expanding.

\paragraph*{HD 45677 }

From their UV - optical low resolution spectropolarimetry
Schulte-Ladbeck \ea\, (1992) infer that HD 45677 is surrounded by a
bipolar nebula. After correcting for ISP, the polarization angle of
the blue/UV spectrum they present is rotated by about 90\degree\, with
respect to the red part of the spectrum.  The  explanation
advanced for this  behaviour is that the red emission is scattered through a
dust torus, while the blue emission, to which the dust is optically
thick, comes from a scattering bipolar flow, perpendicularly oriented
with respect to the torus.

\begin{figure}
\mbox{\epsfxsize=0.48\textwidth\epsfbox{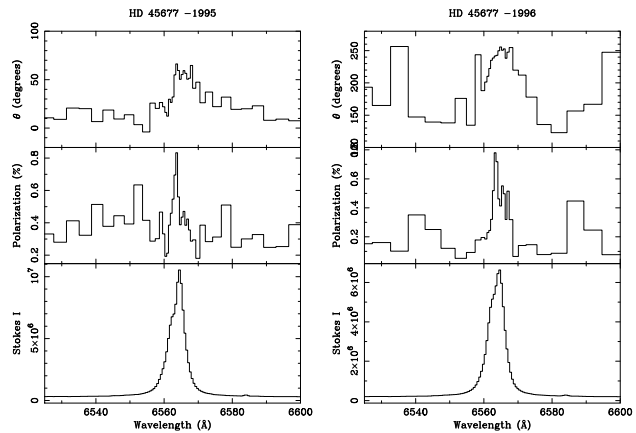}}
\caption{The polarization spectrum of HD 45677  in 1995 and 1996. 
\label{hispec}
}
\end{figure}

The spectra taken in January 1995 and December 1996 are shown in
Fig.~\ref{hispec}. On both occasions the polarization across \ha\ is
enhanced with respect to the continuum.  The variability of the
continuum polarization is very strong, with the polarization changing
from 0.4\% to 0.2\%, while the position angle changed from 11\degree \
to 140\degree .  The peak of \ha\ shows in both cases
roughly the same polarization and position angle ($\sim$0.8\%,
75\degree ). Although \ha\ is enhanced in the observed polarization
spectrum, this does not necessarily imply that the intrinsic spectrum
exhibits the same effect.  In the following we discuss the different
elements, ISP, circumstellar dust scattering and electron-scattering
that shape the observed polarization spectra.  As explained below, we
assume that the polarization changes across \ha \ are due to
electron-scattering.

\begin{figure}
\mbox{\epsfxsize=0.4\textwidth\epsfbox{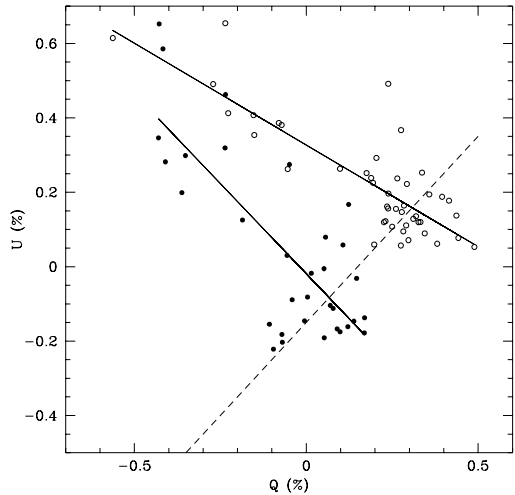}}
\caption{
QU data of HD 45677 of both epochs at 0.04\% binning.  Open
circles represent the January 1995 data, and  filled circles the
December 1996 data. The solid lines are fits through the line data,
and the dashed line represents a fit through the two continuum points.
\label{qu}
}
\end{figure}

The spectra are plotted in QU space in Fig.~\ref{qu}.  Both
circumstellar dust polarization and ISP add only a constant QU vector
to all points.  The continuum points cluster at different positions on
the dates of observation, while the QU vectors across \ha\ are almost
parallel. A least squares fit through the QU-points between 6550 and
6570 \ang , returns intrinsic polarization angles of 168 $\pm$
3\degree \ and 163 $\pm$ 3\degree \ for the 1995 and 1996 spectra
respectively.  Depending on the quadrant where the intrinsic QU
vectors are located, these values could be rotated by 90\degree \ and
define a projected angle on the sky of $\approx$ 75\degree.  The
length of the vector on both occasions corresponds to P $\approx$
0.95\%, assuming the line to be unpolarized. The electron-scattering
region is thus aspherical, has a PA of $\sim$ 75\degree\ on the sky at
a relatively constant amplitude of about 0.95\% .

   How does this relate to the view taken by Schulte-Ladbeck \ea\ (1992) 
of their similar detection of enhanced \ha\ linear polarization?  In their
ISP-corrected spectrum, \ha\ is still enhanced with respect to its
adjacent continuum.  They explained this in terms of an ionized region
much closer to the circumstellar dust than the stellar point source: the
\ha\ line then sees a larger solid angle of scattering material and
is thus more polarized than the continuum.  This seems to us an unlikely
alternative to the conventional view that the ionized region, with a
temperature of $\sim$10000~K, is located within a very much smaller volume
around the star than the dust, which should have an equilibrium 
temperature below $\sim$1500~K, the dust condensation temperature (see
e.g. the spectral energy distribution modelling of HD~45677 by Sorrell
1989).  Furthermore, the ISP correction adopted by Schulte-Ladbeck \ea\
(1992) was the {\em ad hoc} value proposed by Coyne \& Vrba (1976) on the
basis that it should be comparable with the relatively steady observed 
polarization in the blue (the red is much more variable).  Whatever this
correction actually does correspond to, there is no reason to suppose that
it accounts for both the ISP and circumstellar dust polarization.  If both 
sources of foreground polarization can be removed with confidence, only 
then can it be discerned whether there is an intrinsic linear polarization
enhancement across \ha .  Hence, for the timebeing, we retain the more
conventional view that the observed polarization change across \ha\ is
due to electron scattering in the ionized region, which may project as
an oblate `disk' or as a prolate `bipolar flow'.

Now we turn to consider the change in polarization of the continuum
points.  As it is likely that the ISP is constant in time, the
additional, potentially variable, mechanisms involved are
electron-scattering and circumstellar dust scattering.  In the same
manner as the polarization changes across \ha\ move along a constant
angle, temporal changes, due to a polarization mechanism that becomes
stronger or weaker, may only affect the magnitude of polarization and
not the PA.  If we adopt this principle here, the intrinsic PA of the
polarizing material responsible for the change in continuum
polarization is then $\sim$ 20\degree .

Since the intrinsic PA of the electron scattering region is $\sim$
75\degree, and its effect on the \ha\ line appears not to have
changed significantly, the variable component does not seem to
be electron scattering.  This leaves circumstellar dust scattering
as the likely variable.  Although the polarization variability of 
HD 45677 is well known,
dating back to the paper in 1976 by Coyne \& Vrba, it is only the high
spectral resolution of the current data that allows us to discriminate
between  electron and dust scattering.

The major question remaining is why the apparent rotation of about
55\degree\ between the major axis of the (variable) dust polarization
and the \ha\ line forming region exists.  In principle, this would
suggest that multiple geometries, such as a combination of an equatorial
disk and a bipolar flow, would show a rotation of 90\degree , regardless
of the respective opening angles.  The cancellation of
perpendicularly-oriented polarization vectors tends to increase for
larger opening angles, decreasing the observed polarization percentage,
but the 90\degree\ rotation will remain intact.  An answer may lie in
the clumpiness of the dusty material around the object.  HD 45677 is
photometrically variable, a property attributed to the presence of
various dust-clouds surrounding and even orbiting the object (de Winter
\& van den Ancker 1997).  Clumpy material is also revealed by the
spectroscopic variability.  Grady \ea\ (1993) show the presence of
variable redshifted absorption lines which are attributed to infalling
and evaporating cometary bodies.  These indicate patchiness of the
circumstellar material as well. 

The result of such clumpy material on the polarization angle is
relatively easy to understand.  If (one of) the scattering regions is
clumpy, a rotation of 90\degree\ is only retrieved if the clumps are
symmetrically distributed around the central star. If not, not all
perpendicularly oriented polarization vectors will cancel out, and the
net effect is that the observed position angle does not represent the
time-averaged mean orientation of the scattering material.  A similar
argument has been brought forward by Trammell \ea\ (1994) to explain the
rotation of 70\degree\ (instead of 90\degree) in the spectrum of IRAS
08005-2356.  It seems thus that the rotation between the dust ring and the
\ha\ line forming region, which is less than 90\degree , could be the
result of scattering with incomplete cancellation in an inhomogeneous
region.  

A word of caution should be given here with regard to the
`true' orientation of the dusty material. The 20\degree \ that was
measured between the continuum points in Fig.~\ref{qu}, and its
comparison with the intrinsic angle of the electron scattering region
points to the clumpiness.  However, this direction along which the
variation occurred should not now be associated with the orientation
of the scattering region, since we only measure incomplete cancellation
of dust clumps at changing positions.  Only sequences of observations 
of this type may give a clue as to whether the variable dust component 
arises from a region perpendicular to or colinear with the electron 
scattering region.

The main result of the new observations of HD 45677 is that it is
possible to discriminate between the {\it electron} scattering and the
{\it dust} scattering regions.  The former can be probed by the change
across \ha\ while the latter is probed by the variability of the
polarization of the continuum.

\section{Discussion}

\subsection{What the observations tell us}

This paper concerned medium resolution spectropolarimetry of a
relatively large sample of B[e] and Herbig Be stars, objects which so
far have never been observed in this way.  We have described the
results of these exploratory observations in a qualitative way and we
summarize the highlights of both non-detections and detections below,
and then briefly discuss the implications of the results.

\subsubsection{`Non'-detections}

   The main goal of the observations was to answer the basic question 
``Does the ionized material around these stars project to circular symmetry
on the sky, or not?'' by investigating whether or not a `line-effect'
is seen across \ha .  In principle, a non-detection should imply a 
circular projection.  This encompasses three-dimensional geometries that 
are spherically-symmetric, or disk-like seen close to face-on.  

Two  of the non-detections failed in quite different ways to 
fit into this simple picture:

    First, in the case of MWC~297 -- a young B1.5 star in the Aquila
Rift (Drew \ea\ 1997) -- observation over two hours or so produced
evidence of, at most, a subtle line effect.  This was despite the fact
that the radio image of this object shows a clearly elongated ionized
gas distribution.  An intriguing interpretation, testable by high
resolution imaging, is that we view MWC~297 only indirectly at optical
wavelengths.  If the direct sightline to the disk should reveal an
edge-on structure, but if the disk is  obscured from view due to
the large extinction, scattering dust clouds may see a more circularly
symmetric structure, such as a face-on disk, and reflect a
polarization spectrum without a line effect to the observer.

    Second, we have presented data on $\omega$~Ori, a star that has
been reported twice before in the literature as showing an \ha\ line
effect (in both instances the observations were narrow-band rather
than spectropolarimetric).  The absence of any such effect in our data
suggests that the ionized envelope is smaller than at
previous occasions because the,  optically thin, electron-scattering
is more sensitive to changes in ionization than the optically thick
\ha \ emission.  The non-detection in $\omega$ Ori implies that single
epoch measurements are not always sufficient to provide a definitive
answer on the circumstellar geometry of these objects.

    In both cases, it is clear that \ha\ spectropolarimetry is best
judged in the context of other observational constraints on the target.  
Indeed they warn not to assume {\em too} quickly that the ionized regions 
in stars without a line effect are face-on disks or spherically-symmetric.

\subsubsection{Objects displaying  a line-effect}

In contrast, the detection of a line-effect immediately tells us that
the scattering region does not project to circular symmetry on the
sky.  The data presented in this paper provide a new, richer variety
of line effects than has hitherto been seen in the literature.  This
is in part due to their relatively high spectral resolution.

The curious cases of the somewhat similar situations of HD 37806 and
HD 50138 offer great opportunities to understand the conditions close
to the star.  Both objects show double peaked \ha \ line profiles,
that have different $V/R$ ratios in normal intensity spectra, but
which turn out to be of equal strength in the polarized flux data.
Both stars also exhibit a superposed single component of \ha\ emission
that is also picked out by a change in linear polarization at much the
same wavelength.  This suggests that the line profile as a whole may
be the result of two kinematically-distinct phenomena.  These may be a
rotating disk (or self-absorbed compact nebula), and a spatially more
extended region of less-rapidly expanding H{\sc ii} whose emission is
only polarized by the ISM.

    A particularly striking result to emerge from our data concerns
the probable Herbig Be star, HD 53367.  We have observed
depolarization across \ha\ without angle rotation.  The co-alignment
of the local interstellar magnetic field and stellar rotation axis
together with the findings of Vrba \ea \ (1987), favour formation of
this relatively massive star ($M > 10$~M$_{\odot}$) by collapse rather
than by the merger of less massive stars.

    The power of repeated observations is shown by the case of HD 45677.
Due to the continuum variability and the constant polarization arising 
from electron scattering, it is possible to distinguish between the
electron and circumstellar dust scattering mechanisms.  This is the first 
time that it has been possible to do this. Since the measured intrinsic 
angles of the dusty and the ionized region differ by 55\degree \ rather 
than 90\degree \ (for perpendicular geometries) or 0\degree \ (for parallel 
geometries) we can conclude that the the dust component is clumpy.

\subsection{Implications for Herbig Be and B[e] star research}

Among the probable Herbig~Be stars we have observed, around half of
them have shown no detectable line effect (Lk\ha ~218, HD~52721,
V380~Ori and MWC~297).  This of course means that almost half have
(namely HD~259431, HD~37806 and HD~53367) and should encourage further
campaigns of this nature.  Using narrow bands, Poeckert \& Marlborough
(1976) surveyed 48 Be stars and found that 21 showed the line effect
at a 3$\sigma$ level, while a further 8 show the line effect at a
2-3$\sigma$ level.  They investigated the relation between the
intrinsic polarization of these Be stars (measured from the \ha\
polarization change) with $v \sin i$ (as measure of inclination) and
found that their observations could well be explained by inclination
effects.  Although based on very small number statistics, the
comparable incidence of line effects in the Herbig Be stars observed
does hint that flattened ionized circumstellar structures are quite
common for this object class as well, and that the non-detections in
 our sample could be due to random sampling of the full range of
inclinations.  One of our non-detections is of course MWC~297, an
object revealed by radio imaging to be non-circular (albeit on a scale
of tens of AU).

     It is important to appreciate that the \ha\ polarization effect
is sensitive to much smaller structures than are presently probed
directly by imaging. Analytical calculations such as those by by
Cassinelli, Nordsieck \& Murison (1987, their Fig. 7) demonstrate that
the bulk of electron scattering occurs on scales as small as two to
three stellar radii. If the Herbig Be stars we have observed are
indeed young or even pre-main sequence objects, then the deviation
from spherical symmetry of the ionized region is presumably a
consequence of the way in which they have formed.  Viewed in these
terms, the structures that we detect now via these polarization
measurements could well be accretion disks that reach to within a
stellar radius or so of the stellar surface.  This conclusion is at
variance with the magnetospheric accretion model widely regarded as
applicable to T~Tau stars, wherein magnetic channeling inhibits the
formation of the inner disk (see Shu \ea \ 1994).  For Herbig Be stars
it remains a more open question as to how far magnetic fields
determine the accretion geometry.  Since the main sequence destiny of
these more massive stars is to possess radiative envelopes, it leaves
more room to doubt that magnetic fields must play a big role at the
stage in which we are able to observe them.  It will be interesting to
see if further \ha\ spectropolarimetry continues to uncover plausible
disk accretors.

    A conceptual model of how these objects may look, still embedded 
in accretion disks reaching into the stellar surface, has recently been 
devised by Drew, Proga \& Stone (1998, building on the work of Proga, 
Stone \& Drew 1998).  This work shows how observationally-significant 
disk winds, driven by radiation pressure, would be created.  A further 
piece in the puzzle of Herbig Be and BN objects that these predicted flows can 
help explain is the high contrast, quite narrow H{\sc i} line emission 
often observed at earlier B spectral types (Drew, 1998).

     A further very strong outcome of this study is that all objects
that can be classified as (evolved) B[e] stars, presented significant
polarization changes across \ha .  A factor that clearly helps
increase the likelihood of detecting a spectropolarimetric line effect
in B[e] stars is that their \ha\ profiles are typically extremely high
contrast and often somewhat broader than in Herbig~Be stars.
Pre-eminent among our B[e] group is HD~87643 which has already been
discussed in a separate paper (Oudmaijer et al. 1998).  Here we have
presented MWC~137 (probably an evolved B[e] star), HD~45677 and
HD~50138.  The fact that {\it all} observed Galactic B[e] stars in our
sample show the line-effect in one incarnation or another lends strong
support to the Zickgraf \ea \ (1985, 1986) model. The variety of
line-effects observed in our data illustrate that the structures
around these stars have their deviation from spherical symmetry in
common, but that the details in each  case are different.
So far, the discussion of these data has been largely qualitative.  As
models simulating these phenomena begin to be calculated there will no
doubt be a considerable sharpening of insight.

\section{Final remarks}

     Apart from providing some striking insights into a number of the
targets observed, the programme of observations we have described here
has offered some lessons in how best to obtain single-line
spectropolarimetric data.  It is clear that the spectral resolution
available to us ($R \approx 5000$) has in most instances been just
enough. As numerical modelling becomes more commonplace, the case
for increased spectral resolution will become stronger.  The main
issue, nevertheless, is the achievement of high enough data quality.
Our 8$^{th}$ magnitude and brighter objects have come out well in
under an hour's telescope time, while 10$^{th}$ to 12$^{th}$ magnitude
objects require several hours observation with a 4-metre class
telescope in at least middling weather conditions.  Ultimately, these
fainter sources will be best served by 8-metre class facilities where
the shorter total integration times will be less subject to weather
influence -- presently they can be an uncertain struggle. 

     The overall conclusion of this study is that this relatively
unexplored mode of observing does yield valuable new insights.  In
some instances we have encountered deepening mysteries that suggest
conclusions drawn from other data have missed something.  MWC~297 and
HD~45677 are both good examples of this.  At the same time, \ha\
spectropolarimetry readily throws up examples that demand
sophisticated numerical modelling of a type that is just beginning to
become available (Hillier, 1996; Harries, 1996).

\paragraph*{\it Acknowledgments}

We thank the staff at the Anglo-Australian Telescope for their expert
advice and support.  Conor Nixon and Graeme Busfield are thanked for
their help during some of the observing runs.  The allocation of time
on the Anglo-Australian Telescope was awarded by PATT, the United
Kingdom allocation panel.  RDO is funded by the Particle Physics and
Astronomy Research Council of the United Kingdom.  The data analysis
facilities are provided by the Starlink Project, which is run by CCLRC
on behalf of PPARC.  Part of the observations are based on data
obtained from the William Herschel Telescope, Tenerife, Spain, in the
Isaac Newton Group service scheme.  This research has made use of the
Simbad database, operated at CDS, Strasbourg, France.

\end{document}